\documentclass[a4paper,11pt]{article}
\usepackage{jheppub} % for details on the use of the package, please
\usepackage[T1]{fontenc} % if needed
\usepackage{amsmath, amssymb, amsfonts, latexsym, mathtools}
\usepackage{physics}
\usepackage{bm}
\usepackage{cancel}
\usepackage{color}
\usepackage{simplewick}

\usepackage{tikz-feynhand}
\usepackage{lscape}

\title{\boldmath Non-invertible Selection Rules from Generalized Discrete Gauging of Finite Non-Abelian Symmetries}

\author[a]{Hiroshi Ohki}
\author[b]{Shohei Uemura}

\affiliation[a]{Department of Physics, Nara Women's University, Nara 630-8506, Japan}
\affiliation[b]{Faculty of Education, Nara University of Education, Nara 630-8528, Japan}

\emailAdd{hohki@cc.nara-wu.ac.jp}
\emailAdd{uemura.shohei.ry@cc.nara-edu.ac.jp }

\abstract{
We investigate non-invertible selection rules originating from the discrete $H$-gauging of theories 
with an underlying discrete global symmetry group $G$. 
To systematically describe these theories, we formulate a general framework 
for $H$-gauged models that incorporates generalized field transformations. 
Our approach naturally accommodates non-Abelian groups, 
for which multidimensional irreducible representations play an essential role. 
In such models with non-Abelian groups, the transformations induced by $H$ non-trivially mix the internal components of $G$-multiplets, 
potentially projecting out specific degrees of freedom. 
Consequently, conventional selection rules based on standard tensor product decompositions or conjugacy classes become insufficient. 
By analyzing the full semidirect product $G \rtimes H$, we introduce projected characters 
to derive necessary and sufficient conditions for non-vanishing $n$-point bare couplings. 
Furthermore, we demonstrate that the remaining field components obey an associative fusion-like algebra 
governed by their Clebsch-Gordan coefficients. 
Phenomenologically, these selection rules restrict allowed interactions 
and impose specific relations among coupling constants. 
We illustrate our results through concrete examples, 
including $\Delta(54) \cong \Delta(27)\rtimes \mathbb{Z}_2$ and $S_4 \cong A_4 \rtimes \mathbb{Z}_2$.
}

\begin{document} 
\maketitle
\flushbottom

\section{Introduction}

In recent years, generalizations of conventional group symmetries have attracted attention and have been actively studied.
Among these developments, non-invertible symmetry has emerged 
as a novel class of generalized symmetry which can not be described by standard group symmetry%
\footnote{For reviews, see \cite{Cordova:2022ruw, Shao:2023gho, Schafer-Nameki:2023jdn, 
Brennan:2023mmt, Costa:2024wks, Kaidi:2026urc} and references therein.}.
In such frameworks, the symmetry transformations lack inverse elements and therefore do not form a traditional group.
Nevertheless, the corresponding selection rules are governed 
by a more general algebraic or categorical structure, which characterizes these symmetries as non-invertible.

There are a number of examples realizing non-invertible symmetry such as two-dimensional conformal field theory \cite{Verlinde:1988sn, Moore:1988qv, Moore:1989yh, Fuchs:1993et, Fuchs:2002cm, Frohlich:2004ef, Frohlich:2006ch, Bhardwaj:2017xup, Chang:2018iay, Thorngren:2019iar, Thorngren:2021yso}.
In string orbifolds, string states are labeled by the conjugacy classes $[g_i]$ of the orbifold group $G$, 
which characterize the corresponding twisted boundary conditions \cite{Dijkgraaf:1987vp, Kobayashi:2004ya, Dixon:1985jw, Dixon:1986jc, Hamidi:1986vh, Dixon:1986qv, Kobayashi:2025ocp}.
There are non-vanishing $n$-point bare couplings only when there is a group element $\tilde g_i$ in each $[g_i]$ such that $\tilde g_1 \tilde g_2 \cdots \tilde g_n = e$ \cite{Kobayashi:2006wq,  Heckman:2024obe, Lin:2022dhv,  Kaidi:2024wio}.
The resulting selection rule is not described by a conventional group symmetry, 
but rather by a more general algebraic structure associated with the conjugacy classes. It is therefore regarded as a non-invertible selection rule.
Magnetized toroidal orbifolds \cite{Abe:2009vi, Kobayashi:2024yqq, Funakoshi:2024uvy} and Calabi-Yau compactifications \cite{Dong:2025pah} also provide realizations of non-invertible selection rules.
In magnetized orbifolds, interactions are determined by overlap integrals of wave functions in the extra dimensions. 
If the extra dimensions admit geometric symmetries such as isometries, the wave functions transform under the corresponding geometric transformations, leading to selection rules for the interactions. 
When such a geometric symmetry is broken by orbifolding, the symmetry is no longer realized as an ordinary invertible symmetry, while the associated selection rules can remain. 
These selection rules are then interpreted in terms of a non-invertible symmetry.

In this paper we investigate non-invertible selection rules arising from discrete gauging of global symmetry.
We start with a four-dimensional quantum field theory having a global discrete symmetry $G$.
A field $\phi_\alpha$ in this model is labeled by an irreducible representation $\mathbf{r}^{(\alpha)}$ of $G$.
Next, we consider a group of automorphisms $H$ acting on $G$, which can be associated with orbifolding in extra dimensions.
The full symmetry of the theory is thereby enhanced to the semidirect product $G \rtimes H$.
$H$-gauging is realized by restricting the physical states to $H$-invariant subspace.
If $H$ and $G$ do not commute with each other, the $H$-invariant state is not preserved under $G$, 
and the original $G$-symmetry is broken.
However, a remnant of this symmetry survives, giving rise to non-invertible selection rules.
Such constructions have been explored in recent literature~\cite{Kobayashi:2024cvp, Dong:2025jra}%
\footnote{Gauging a normal subgroup of the full symmetric models is also considered in \cite{Tachikawa:2017gyf}. 
$G$-gauging of non-Abelian $G$-symmetric models is investigated in \cite{Chang:2018iay, Heidenreich:2021xpr, Bhardwaj:2022yxj}.}.
In previous works, $G$ has been restricted to be Abelian.
In these cases, $\mathbb{Z}_N$ charges play an important role in the selection rules.
For example, in the case of $\mathbb{Z}_2$ gauging of $\mathbb{Z}_N$ symmetric models, 
$\mathbb{Z}_2$-invariant fields carry both charges of $q$ and $-q$ simultaneously, 
which lead to a selection rule for $n$-point interaction of the form $q_1 \pm q_2 \pm \cdots \pm q_n = 0$.
In addition, Abelian groups have a manifest correspondence between the charges and group elements, 
and the selection rule can be reinterpreted as a non-invertible selection rule based on conjugacy classes.
Phenomenologically, these non-invertible selection rules provide a novel approach to model building; for instance, 
they can generate useful Yukawa matrix textures and prohibit the strong CP phase or other dangerous terms~\cite{Kobayashi:2024cvp, Suzuki:2025oov, 
Chen:2025awz, Okada:2025kfm, Nakai:2025thw, Jangid:2025krp, Jiang:2025psz, Liang:2025dkm, Nomura:2025tvz, 
Nomura:2025yoa, Kobayashi:2025cwx, Kobayashi:2025wty, Kobayashi:2025thd, Kobayashi:2025rpx, Kobayashi:2025znw, 
Nomura:2026hli, Dong:2026iwa, Okada:2026gxl, Qu:2026omn, Okada:2026bpp, Okada:2026pek, Chen:2026mvi, 
Nomura:2026hcu, Kitagawa:2026eck, Kobayashi:2026msq}.

On the other hand, when $G$ is non-Abelian, such a clear one-to-one correspondence does not exist, 
yet we find that analogous non-invertible selection rules still apply.
An automorphism $h$ maps an irreducible representation $\rho^{(\alpha)}$ to $ \rho^{(h\cdot \alpha)} \equiv \rho^{(\alpha)}\circ h$.
Thus, the $H$-invariant field originating from $\mathbf{r}^{(\alpha)}$ contains a combination of representations $\{\mathbf{r}^{(\alpha)}, \mathbf{r}^{({h_1} \cdot \alpha)}, \mathbf{r}^{({h_2} \cdot \alpha)}, \cdots\}$.
The necessary condition for non-vanishing $n$-point bare interaction $\tilde \phi_{\alpha_1} \tilde \phi_{\alpha_2} \cdots \tilde \phi_{\alpha_n}$ 
is the existence of $h_i\in H$ for each $\alpha_i$ which satisfies
\begin{align}
	\mathbf{r}^{({h_1}\cdot\alpha_1)} \otimes \mathbf{r}^{({h_2}\cdot\alpha_2)} \otimes \cdots \otimes \mathbf{r}^{({h_n}\cdot\alpha_n)} \ni \mathbf{1}.
\end{align}
However, this selection rule is not complete, 
since the multidimensional $H$-transformations eliminate some components of the multiplet fields, 
which can project out the trivial singlet appearing in standard tensor products.
As a result, the above selection rule is necessary but not sufficient.

To obtain a complete selection rule, 
we must construct a quantity that properly tracks the remaining degrees of freedom after the gauging procedure.
We achieve this by analyzing the representations of the full semidirect product $G\rtimes H$ rather than just $G$. 
This approach enables us to find a general formula that is valid even for generally non-Abelian $H$-gauging. 
We demonstrate that the orthogonality relations of these irreducible representations still provide a rigorous way to count the remaining singlet components in the tensor product after $H$-gauging. 
Specifically, we introduce a projected character $\tilde \chi^{(\alpha)}$, defined as 
the projected trace of representation onto the \(H\)-invariant subspace.
Using this, the necessary and sufficient condition for an allowed $n$-point bare coupling is given by 
\begin{align}
	\sum_{g \in G \rtimes H} \tilde \chi^{(\alpha_1)}(g) \tilde \chi^{(\alpha_2)}(g) \cdots \tilde \chi^{(\alpha_n)}(g) \neq 0.
\end{align}
Furthermore, we explore the algebraic structure of these remaining fields. 
While their vector spaces do not form a closed algebra under standard tensor products, 
their individual components obey an associative fusion-like algebra governed by their Clebsch-Gordan (CG) coefficients. 
Phenomenologically, this shows a clear difference from Abelian models: 
these non-invertible selection rules not only restrict allowed bare couplings but also impose specific relations among the coupling constants of multiplet interactions. 
These constraints offer an interesting framework for particle phenomenology, particularly in flavor model building.

This paper is organized as follows.
In section \ref{sec:review}, we briefly review 
$H$-gauging of theories with $G$ symmetry and fix our terminology. 
We first review $H$-gauging of abelian symmetries, then we consider $H$-gauging of non-Abelian symmetries.
We find naive selection rule is not sufficient for non-Abelian symmetric models, and consider why not.
In section \ref{sec:rep} , we study $H$-gauging of $G\rtimes H$ symmetric models. 
We construct associative multiplication rules for $H$-invariant fields.
We use CG coefficients to define the multiplication rule.
Then we rewrite the non-invertible selection rules by characters of irreducible representations restricted to $H$-invariant subspace.
In section \ref{sec:exmple}, we construct some explicit examples.
In section \ref{sec:pheno}, we study phenomenological implications.
We find when $G$ is non-Abelian, non-invertible selection rules may restrict flavor structures of Yukawa matrices.
Section \ref{sec:conclusion} is devoted to the conclusion.
In Appendix \ref{sec:tilde_rho}, we show orthogonality relations for projected characters.
In Appendix \ref{sec:CG_D54}, we summarize the CG coefficients for $\mathbb{Z}_2$-invariant representations of $\Delta (54)$.

\section{Generalization of Discrete Gauging by Automorphisms}
\label{sec:review}

Discrete global symmetries play an important role in particle phenomenology.
These symmetries are occasionally gauged by another discrete group 
via a group action such as an orbifold projection in a higher-dimensional theory.
Following an orbifold projection, physical states are restricted to the orbifold-invariant states, 
which is equivalent to discrete gauging.
While the original global symmetry is generally no longer realized as an ordinary global symmetry,
a remnant non-invertible selection rule may remain.
In this section, 
we first extend the $H$-gauging of the symmetry group $G$ 
by generalizing the field transformations under $H$.
We note that, in order to determine the full symmetry properties of the theory, 
we must take into account the field transformations under $H$ as well as its action on $G$.
Namely, we should consider the symmetry structure associated with the group generated by $G$ and $H$, 
denoted by $\langle H,G\rangle$, where the field transformation matrix $U$ should satisfy the consistency condition.
In this case, bare couplings are allowed when the $H$-invariant fields contain 
appropriate representations whose tensor product contains the trivial singlet.
We find, however, that this criterion provides only a necessary condition for the existence of such couplings and is not sufficient in general.

\subsection{Generalized Field Transformations in Discrete Gauging}
\label{sec:Generalized Field Transformations}

Let us consider a theory with a global symmetry $G$ and its discrete gauging by an automorphism group $H$ of $G$. 
We assume a Lagrangian containing scalar fields $\phi_\alpha$, which are labeled by their irreducible representations $\mathbf{r}^{(\alpha)}$ under the symmetry group $G$. 
For simplicity, we restrict our discussion to scalar fields, as the generalization to fermionic fields is straightforward.

An automorphism $h \in H$ induces a mapping from $\mathbf{r}^{(\alpha)}$ to $\mathbf{r}^{(\beta)}$ such that $\rho^{(\alpha)} \circ h \sim \rho^{(\beta)}$.
Let $\bar{h}$ denote the linear map induced by the automorphism $h$
. 
The transformation of $\phi_\alpha$ under $\bar{h}$ is then given by
\begin{align}
	\phi_\alpha \to \bar{h}\phi_\alpha = U \phi_\beta,
	\label{eq:action_h}
\end{align}
where the unitary matrix $U$ must satisfy the consistency condition
\begin{align}
	\rho^{(\alpha)}(h(g)) = U^{\dagger} \rho^{(\beta)}(g) U.
	\label{eq:auto_matrix}
\end{align}
Here, $U$ maps the representation space $V^{(\alpha)}$ to $V^{(\beta)}$. 
This is the standard consistency condition for the transformation of a representation of $G$ induced by an automorphism $h$ (see, e.g., \cite{Holthausen:2012dk, Chen:2014tpa, Ohki:2023zsn} for generalized CP).

Suppose that $U$ and $V$ are both solutions to Eq.~\eqref{eq:auto_matrix}. 
Then we find $V^\dagger \rho^{( \beta)}(g) V = U^{\dagger}\rho^{(\beta)}(g) U $,
which implies that $\rho^{(\beta)}(g)$ and $UV^{\dagger}$ commute for all $g \in G$. 
For an irreducible representation $\mathbf{r}^{(\beta)}$, 
Schur's lemma implies that $UV^\dagger$ must be proportional to the identity matrix. 
Hence, we have $V = e^{i\theta}U$, meaning that $U$ is unique up to an overall phase.
However, the relative phases among transformations associated with different elements of $H$ 
are subject to further constraints arising from the consistency conditions of the group extension by $G$. 
Thus, specifying the full transformation of a field requires not only its irreducible representation under $G$, but also the explicit $H$-transformations on the corresponding representation space.

Let $h_1,h_2\in H$ and suppose that $U_{h_1}$ maps $V^{(\alpha)}$ to $V^{(\beta)}$, while $U_{h_2}$ maps $V^{(\beta)}$ to $V^{(\gamma)}$. 
Then the product $U_{h_2}U_{h_1}$ maps $V^{(\alpha)}$ to $V^{(\gamma)}$. 
On the other hand, $U_{h_2h_1}$ also maps $V^{(\alpha)}$ to $V^{(\gamma)}$. 
Therefore, we obtain the following consistency condition:
\begin{align}
(U_{h_2}U_{h_1})^\dagger
\rho^{(\gamma)}(g)
(U_{h_2}U_{h_1})
&=
U_{h_1}^\dagger
\left[
U_{h_2}^\dagger
\rho^{(\gamma)}(g)
U_{h_2}
\right]
U_{h_1}
\notag\\
&=
\rho^{(\alpha)}
\bigl((h_1h_2)(g)\bigr),
\end{align}
which coincides with the transformation induced by $U_{h_2 h_1}$.
Therefore, the operator
\begin{align}
U_{h_2}U_{h_1}U_{h_2h_1}^\dagger
\end{align}
commutes with $\rho^{(\gamma)}(g)$ for all $g\in G$. 
The usual consistency condition implicitly requires this operator to be the identity,
\begin{align}
U_{h_2}U_{h_1}
=
U_{h_2h_1}.
\label{eq:trivial_cocycle}
\end{align}
In this case, the transformations of $G$ and $H$ on the representation spaces combine to give a representation of the semidirect product $G\rtimes H$.

This condition, however, is more restrictive than necessary. 
For example, when $G$  and $H$ are Abelian groups, 
since $U_{h_2}U_{h_1}U_{h_2h_1}^\dagger$ commutes with the representation of $G$, Schur's lemma allows us to introduce a generalized relation
\begin{align}
U_{h_2}U_{h_1}
=
\rho^{(\gamma)}
\bigl(f(h_1,h_2)\bigr)
U_{h_2h_1},
\qquad
f(h_1,h_2)\in G.
\label{eq:cocycle}
\end{align}
Mathematically, it is well known that the function $f: H\times H \to G$ must satisfy the 2-cocycle condition to ensure the associativity of the operators $U_h$. 
In the language of group cohomology, such inequivalent group extensions are classified by the second cohomology group $H^2(H, G)$. 
Relaxing the initial restriction allows us 
to consider these more general group extensions\footnote{For review of group extension see a mathematical textbook such as \cite{Serre:2016fg}.}.

For example, the automorphism group of $\mathbb{Z}_4$ is isomorphic to $\mathbb{Z}_2$. 
Let $h$ denote its generator, which acts on the generator $a$ of $\mathbb{Z}_4$ as
\begin{align}
h: a \to a^3=a^{-1}.
\end{align}
There are two inequivalent choices for the lift of $h$ to the corresponding group extension, characterized by
\begin{align}
h^2=e
\qquad\text{or}\qquad
h^2=a^2.
\end{align}
The first choice gives the dihedral group $D_4$, while the second gives the quaternion group $Q_8$. 
Both are extensions of $\mathbb{Z}_2$ by $\mathbb{Z}_4$ with the same induced outer action.

In this paper, we restrict ourselves to the trivial cocycle,
\begin{align}
f(h_1,h_2)=e
\qquad
\text{for all }h_1,h_2\in H,
\end{align}
so that the resulting group extension is the semidirect product $G\rtimes H$. 
As a direct consequence of this restriction, if $H$ is a cyclic subgroup,
\begin{align}
H=\langle h \rangle\simeq\mathbb{Z}_m,
\end{align}
the corresponding unitary matrix satisfies $\bar{h}^m=1$, and thus $U^m=1$. 
We will study the $H$-gauging of $G$ whose transformation is consistent with the semidirect product of $G \rtimes H$ in the following subsection, leaving the case of nontrivial cocycles, such as the $Q_8$ extension above, for future work.

\subsubsection*{Discrete $H$-gauging}

We next consider an $H$-gauging, where $H$ is a subgroup of the automorphism group of $G$. 
The $H$-gauging is defined by restricting the physical fields to $H$-invariant combinations. 
Equivalently, fields related by the action of $H$ are identified. 
The group $H$ acts on the irreducible representations of $G$ according to
\begin{align}
\mathbf{r}^{(\alpha)}
\longrightarrow
h\cdot\mathbf{r}^{(\alpha)},
\qquad
h\in H,
\end{align}
where $h\cdot\mathbf{r}^{(\alpha)}$ is defined by $\rho^{(\alpha)}\circ h \simeq \rho^{(h\cdot\alpha)}$. 
Let $H_\alpha$ denote the stabilizer of $\mathbf{r}^{(\alpha)}$ under this action:
\begin{align}
H_\alpha
=
\left\{
h\in H
\ \middle|\ 
h\cdot\mathbf{r}^{(\alpha)}
\simeq
\mathbf{r}^{(\alpha)}
\right\}.
\end{align}
The orbit-stabilizer theorem implies that there is a one-to-one correspondence between the orbit of $\mathbf{r}^{(\alpha)}$ under $H$ and the left coset space $H/H_\alpha$. 
The $H$-invariant fields $\tilde{\phi}_\alpha$ are then generally given by 
\begin{align}\label{eq:H-invariant}
\tilde{\phi}_\alpha =& \frac{1}{|H/H_\alpha|} \sum_{h_i \in H/H_\alpha} \bar{h}_i P_H^{(\alpha)} \phi_\alpha, 
\end{align}
where $h_i$ are representatives of the left cosets, and $P^{(\alpha)}$ is a projection operator from $V^{(\alpha)}$ to the subspace of $H_\alpha$-invariant vectors,
\begin{align}\label{eq:P}
P_H^{(\alpha)} =& \frac{1}{|H_\alpha|} \sum_{h \in H_\alpha} U_h.
\end{align}
From the viewpoint of orbifold projections, $\phi_\alpha$ is projected onto the $H$-invariant combination $\tilde{\phi}_\alpha$, 
while the remaining linearly independent combinations are projected out.

When $G$ is Abelian, all of its irreducible representations are one-dimensional (singlets). 
Consequently, the unitary transformation matrix reduces to a complex phase, 
$U_h = e^{i\theta_h}$, and the projection operator $P_H^{(\alpha)}$ reduces to either $1$ or $0$. 
The $H$-invariant field remains in the spectrum only when $P_H^{(\alpha)}=1$.  
On the other hand, when $G$ is non-Abelian, the irreducible representations are generally multidimensional, 
and a non-trivial subspace of $V^{(\alpha)}$ may remain after the projection, 
depending on the unitary matrix $U_h$ on the representation space.

These features can be understood more naturally from the viewpoint of the irreducible representations of the semidirect product $G\rtimes H$. 
Different irreducible representations of $G\rtimes H$ can reduce to the same irreducible representation of $G$ upon restriction to $G$; the $H$-action distinguishes these representations. 
Thus, the projection operator acts on a single irreducible representation of $G\rtimes H$, which can be defined more simply as will be shown in Eq.~\eqref{eq:P_H}. 
After $H$-gauging, only the $H$-invariant components remain, while the non-singlet components under $H$ are projected out.

\subsection{Example: Discrete gauging of $\mathbb{Z}_{N}$ model}
\label{sec:gauging_of_Z_N}

As an example, let us consider an Abelian global symmetry $G=\mathbb{Z}_{N}$ and its $H$-gauging, where $H$ is a subgroup of the automorphism group of $G$. 
The automorphism group of $\mathbb{Z}_N$ is $\mathrm{Aut}(\mathbb{Z}_N)\simeq (\mathbb{Z}_N)^\times$, 
where $(\mathbb{Z}_N)^\times$ denotes the multiplicative group of integers modulo $N$ that are coprime to $N$.

For simplicity, we consider a cyclic subgroup $H=\langle h\rangle = \mathbb{Z}_2$, where $h$ acts on the generator $a \in \mathbb{Z}_N$ as
\begin{align}
h : a\to a^{N-1}=a^{-1}.
\end{align}
The corresponding group extension is the semidirect product $G \rtimes H = D_N$, 
as discussed above\footnote{The following discussions can be straightforwardly extended to other outer automorphism groups of $\mathbb{Z}_{N}$, such as $G \rtimes H = T_N$.}. 
A field $\phi_q$ with charge $q$ under $\mathbb{Z}_N$ transforms as
\begin{align}
\phi_q \to \phi'_q
= \rho^{(q)}(a)\phi_q,
\qquad
\rho^{(q)}(a)=e^{\frac{2\pi i}{N}q}.
\end{align}
The automorphism $h$ also induces an action on the representations of $G$. In particular,
\begin{align}
(\rho^{(q)}\circ h)(a)
=\rho^{(q)}(a^{N-1})
=\rho^{((N-1)q)}(a)
=\rho^{(-q)}(a),
\end{align}
so that the charge $q$ is mapped to $-q$ modulo $N$.
The corresponding transformation of the fields, denoted by $\bar{h}$, acts on the fields as:
\begin{align}
\phi_q\to\bar{h}\phi_q=\phi_{-q}.
\label{eq:ZN_inversion}
\end{align}
This transformation satisfies $\bar{h}^2=1$, consistent with $h^2=1$ 
in the semidirect product\footnote{From the consistency condition, one may also consider the transformation 
$\phi_q\to-\phi_{-q}$ instead of $\phi_q\to\phi_{-q}$. The additional phase can be absorbed by a field redefinition and does not lead to a physically inequivalent representation. 
In fact, the two choices are related by a unitary equivalence in the representations of $D_N$.}.

As previously established, the $H$-gauging is defined by restricting the physical fields to $H$-invariant combinations; 
equivalently, fields related by the action of $H$ are identified. 
In this specific model, the charges are organized into orbits under $H$,
\begin{align}
q \leftrightarrow -q.
\end{align}
Following Eq.~\eqref{eq:H-invariant}, the $H$-invariant field is given by 
\begin{align}\label{eq:tildephiq}
\tilde{\phi}_{q}
=\frac{1}{2}\sum_{i=0}^{1}\bar{h}^{i}\phi_q
=\frac{1}{2}
\left(
\phi_q+\phi_{-q}
\right).
\end{align}
From the viewpoint of the orbifold projection, $\phi_{\pm q}$ are projected onto the $H$-invariant combination $\tilde{\phi}_{q}$, 
while the remaining linearly independent combination $\phi_q -\phi_{-q}$ is projected out.

The original $G$ symmetry is no longer realized as an ordinary global symmetry after the $H$-gauging, since $H$ acts on $G$ as an outer automorphism and $\tilde{\phi}_q$ does not transform in a definite representation of $G$. 
Nevertheless, a non-invertible selection rule associated with the original $G$ symmetry remains, as follows.

\subsubsection*{Selection Rules and Fusion-Like Algebra for $H$-Orbits of Conjugacy Classes}

The original Lagrangian, i.e., without $H$-gauging, is assumed to be invariant under $G$. 
Consequently, an $n$-point interaction $\phi_{q_1}\phi_{q_2}\cdots\phi_{q_n}$ is allowed only if the sum of the charges vanishes modulo $N$:
\begin{align}
q_1+q_2+\cdots+q_n=0
\qquad \mathrm{mod}\ N.
\end{align}
For the original $G$ symmetry, this condition is both necessary and sufficient for a coupling to be allowed.

The effective Lagrangian after $H$-gauging is described in terms of the $H$-invariant fields $\tilde{\phi}_q$ rather than the original fields $\phi_q$. 
Therefore, an $n$-point coupling of the $H$-invariant fields is allowed when there exists a choice of signs such that
\begin{align}
\epsilon_1q_1+\epsilon_2q_2+\cdots+\epsilon_nq_n
=0
\qquad \mathrm{mod}\ N,
\qquad
\epsilon_i=\pm1.
\label{eq:selection_rule}
\end{align}
Thus, the $H$-invariant field $\tilde{\phi}_q$ behaves as if it carries both charges $q$ and $-q$. 
This gives the coupling selection rule associated with the $\mathbb{Z}_2$-gauging of the $\mathbb{Z}_N$ symmetry~\cite{Dong:2025jra}.

This selection rule is not described by a conventional group symmetry, but by a more general algebraic structure, and is therefore regarded as a non-invertible selection rule. 
In fact, the above non-invertible selection rule can be reconstructed from the multiplication rules for $H$-orbits of conjugacy classes. 
A conjugacy class of $\mathbb{Z}_N$ is given by a single element: $C^{(q)} = \{a^q\}$ where $q = 0, 1, \cdots , N-1$. 
The automorphism $h$ maps $C^{(q)}$ to $C^{(-q)}$. 
Thus, $C^{(q)}$ and $C^{(-q)}$ belong to the same orbit under the action of $h$. 
We then define the $H$-orbit of $C^{(q)}$ as 
\begin{align}\label{eq:k-orbit}
\tilde{C}^{(q)} = \{a^q,a^{-q} \},
\end{align}
where $q=0, \cdots, N/2$ for even $N$, and $q=0, \cdots, (N-1)/2$ for odd $N$. 
For $q=0$, and also for $q=N/2$ when $N$ is even, the two elements in Eq.~\eqref{eq:k-orbit} coincide, so that the corresponding orbit contains only a single element. 

We define a multiplication rule for the $H$-orbits by multiplying their elements, 
\begin{align}\notag
\tilde{C}^{(q_1)} \otimes \tilde{C}^{(q_2)} 
=& 
\sum_{\epsilon_1=\pm1}
\sum_{\epsilon_2=\pm1}
a^{\epsilon_1 q_1+ \epsilon_2 q_2}
\\ \label{eq:fusion-like}
=&
\tilde{C}^{(q_1+q_2)} \oplus \tilde{C}^{(q_1-q_2)},
\end{align}
where the right-hand side is understood as a formal sum of $H$-orbits, with possible multiplicities for special values of the charges. 
It is obvious that the above multiplication rule satisfies both associativity and commutativity. 
More generally, the product of $n$ orbits 
\begin{align}
\tilde{C}^{(q_1)}
\otimes
\tilde{C}^{(q_2)}
\otimes\cdots\otimes
\tilde{C}^{(q_n)}
\end{align}
contains the identity class $\tilde{C}^{(0)}=\{e\}$ if and only if there exists a set of integers $(\epsilon_1,\epsilon_2,\ldots,\epsilon_n)$ 
with $\epsilon_i = \pm1$ satisfying Eq.~\eqref{eq:selection_rule}. 
Since the $H$-invariant field $\tilde{\phi}_q$ is naturally associated with the $H$-orbit $\tilde{C}^{(q)}$, 
the non-invertible selection rule is reproduced by the multiplication rule of Eq.~\eqref{eq:fusion-like}.

The multiplication rule in Eq.~\eqref{eq:fusion-like} defines a \textit{fusion-like} algebra. 
Unlike standard fusion algebras where fields are labeled by conjugacy classes of $G$ \cite{Kaidi:2024wio}, we label the fields $\tilde{\phi}_q$ by $H$-orbits of conjugacy classes due to the $H$-gauging. 
Furthermore, since $\phi_q$ and $\phi_{-q}$ form a doublet representation $\mathbf{2}_q$ of $D_N$, labeling fields by $H$-orbits is equivalent to labeling them by $D_N$ representations. 
As we will show later, the $H$-orbit multiplication in Eq.~\eqref{eq:fusion-like} exactly matches the tensor-product decomposition of $D_N$.

This correspondence might naively suggest that fields can always be consistently labeled by the representations of $G$. 
However, this picture fails for general $G$. 
In particular, distinct fields belonging to the same irreducible representation of $G$ can transform differently under $H$. 
When $G$ is non-Abelian, these internal components are not fully projected out due to their multidimensionality. 
Consequently, multiplication rules based solely on $G$-representations are generally not closed, and the selection rules cannot be described simply by $H$-orbits. 
As we will demonstrate in the $A_4$ model, this necessitates the generalized $H$-gauging framework based on $G \rtimes H$.

\subsection{Example: $\mathbb{Z}_2$-gauging of $A_4$ Model}
\label{sec:A_4}

Here, we consider the generalized discrete gauging of non-Abelian symmetries in $A_4$-symmetric models.
$A_4$ is the alternating group of order $12$.
It is isomorphic to $\Delta(12) \cong (\mathbb{Z}_2 \times \mathbb{Z}_2) \rtimes \mathbb{Z}_3$; hence, its elements can be parameterized as $a^i a'^j b^k$, where $a$ and $a'$ are generators of $\mathbb{Z}_2$, and $b$ is a generator of $\mathbb{Z}_3$.
These generators satisfy the relations $b^2 a b = a'$ and $b^2 a' b = aa'$ (alongside $a^2 = a'^2 = b^3 = e$).
The fields in this model are labeled by the irreducible representations $\mathbf{r}^{(\alpha)}$ of $A_4$.
The irreducible representations of $A_4$ consist of three singlets $\mathbf{1}, \mathbf{1}', \mathbf{1}''$ and one triplet $\mathbf{3}$.
Here, $\mathbf{1}$ is the trivial singlet, while $\mathbf{1}'$ and $\mathbf{1}''$ are non-trivial singlets.
Their representation matrices are given by
\begin{align}
	\rho^{(\mathbf{1}')}(b) = e^{\frac{2\pi i}{3}}, 
	&&\rho^{(\mathbf{1}')}(a) = \rho^{(\mathbf{1}')}(a') = 1,
	\notag
	\\ 
	\rho^{(\mathbf{1}'')}(b) = e^{-\frac{2\pi i}{3}}, 
	&&\rho^{(\mathbf{1}'')}(a) = \rho^{(\mathbf{1}'')}(a') = 1.
\end{align}
The representation $\mathbf{3}$ is real, and its representation matrices are given by 
\begin{align}
	\rho^{(\mathbf{3})}(a) = 
	\begin{pmatrix}
		1 & 0 & 0 \\
		0 & -1& 0\\
		0 & 0 & -1
	\end{pmatrix},&&
	\rho^{(\mathbf{3})}(a') = 
	\begin{pmatrix}
		-1 & 0 & 0 \\
		0 & -1& 0\\
		0 & 0 & 1
	\end{pmatrix},
	&&
	\rho^{(\mathbf{3})}(b) = 
	\begin{pmatrix}
		0 & 0 & 1 \\
		1 & 0& 0 \\
		0 & 1 & 0
	\end{pmatrix}.
\end{align}

The outer automorphism group of $A_4$ is isomorphic to $\mathbb{Z}_2$.
Let $h$ denote the generator of this outer automorphism, which transforms the generators of $A_4$ as $h: a \to a$, $h: a' \to aa'$, and $h: b \to b^2$.
Consequently, the representations transformed by $h$ satisfy 
\begin{align}
	\rho^{(\mathbf{1}')}(h(g)) = \rho^{(\mathbf{1}'')}(g),
	&&
	\rho^{(\mathbf{1}'')}(h(g)) = \rho^{(\mathbf{1}')}(g),
	&&
	\rho^{(\mathbf{3})}(h(g)) = U^\dagger \rho^{(\mathbf{3})}(g) U,
\end{align}
where $U$ is a $3\times 3$ unitary matrix satisfying this relation for all $g\in A_4$.
We find two distinct solutions for $U$, denoted as $U=U_\pm$, given by
\begin{align}
	U_\pm = \pm \begin{pmatrix}
	1 & 0 & 0\\
	0 & 0 & 1\\
	0 & 1 & 0\\
	\end{pmatrix}.
\end{align}
This overall sign ambiguity is expected from the generalized field transformations and the consistency conditions for automorphisms, as explained in Sec.~\ref{sec:Generalized Field Transformations}.

The automorphism $h$ induces a linear map between the corresponding representation spaces. 
To specify a concrete model of $H$-gauging, let us fix $U=U_-$.  
Since we are constructing an $H$-gauged model based on the symmetry group $G$, 
the transformations of the fields under $\bar h$ should be uniformly given for each $A_4$ representation as
\begin{align}
\label{eq:A4}
	\phi_{\mathbf{1}'} \leftrightarrow  \phi_{\mathbf{1}''},
	\quad
	\phi_{\mathbf{3}} \to U_- \phi_{\mathbf{3}}
\end{align}
such that $\rho^{(\alpha)}(h(g)) = \bar h^{-1} \circ \rho^{(\beta)}(g) \circ \bar h$, 
where $\mathbf{r}^{(\beta)}$ is the representation obtained from $\mathbf{r}^{(\alpha)}$ by the transformation induced by $h$.

After $\mathbb{Z}_2$-gauging, the physical states are restricted to the $\mathbb{Z}_2$-invariant subspace. 
The effective theory is therefore described by $\mathbb{Z}_2$-invariant fields $\tilde \phi_{\alpha}$, which are generally defined as 
\begin{align}
	\tilde \phi_\alpha = \frac 1 2  \left( \phi_{\alpha} + \bar h \phi_{\alpha}\right).
\end{align}
The remaining $\mathbb{Z}_2$-invariant fields are explicitly given by 
\begin{align}
	\tilde{\phi}_{\mathbf{1}} = \phi_{\mathbf{1}},
	&&
	\tilde \phi_{\mathbf{1}'} = \frac 12 (\phi_{\mathbf{1}'} + \phi_{\mathbf{1}''}),
\end{align}
for the singlets, and 
\begin{align}
	\tilde \phi_{\mathbf{3}} = \frac 12 \left(\phi_{\mathbf{3}} + U_- \phi_{\mathbf{3}} \right)
	= 
	\frac 1{2}\left[I_3+ 
	\begin{pmatrix}
	-1 & 0 & 0\\
	0 & 0 & -1\\
	0 & -1 & 0\\
	\end{pmatrix}
	\right]
	\begin{pmatrix}
	x_1\\
	x_2\\
	x_3
	\end{pmatrix}
	= 
	(\tilde \phi_\mathbf{3})
	\begin{pmatrix}
	0\\
	\frac{1}{\sqrt 2}\\
	-\frac{1}{\sqrt 2}
	\end{pmatrix},
	\label{eq:3_tilde}
\end{align}
for the triplet, where we have parameterized $\phi_{\mathbf{3}} = (x_1, x_2, x_3)^T$. 
Here, $(\tilde \phi_\mathbf{3})$ denotes the non-vanishing component of $\tilde \phi_\mathbf{3}$, which is given by 
\begin{align}
	(\tilde \phi_\mathbf{3}) = \frac 1 {\sqrt 2} (x_2 - x_3).
\end{align}
The orthogonal components are projected out. 
Thus, three classes of independent fundamental fields remain in the effective theory, namely $\tilde{\phi}_{\mathbf{1}}, \tilde{\phi}_{\mathbf{1}'}$, and $\tilde \phi_{\mathbf{3}}$.

We now consider the coupling selection rules for general $n$-point interactions. 
An $n$-point bare coupling $\tilde \phi_{\alpha_1} \tilde \phi_{\alpha_2} \cdots \tilde \phi_{\alpha_n}$ originates from an $A_4$-invariant interaction in the original Lagrangian. 
It is therefore allowed if there exists a set of irreducible representations $\alpha_i'$, each related to $\alpha_i$ by an automorphism $h$, such that 
\begin{align}
	\mathbf{r}^{(\alpha'_1)} \otimes \mathbf{r}^{(\alpha'_2)} \otimes \cdots \otimes \mathbf{r}^{(\alpha'_n)} \ni \mathbf{1}_{A_4}.
	\label{eq:cond_orb}
\end{align}
This is analogous to the selection rule obtained from $H$-gauging in models with Abelian symmetries. 
However, as we will explicitly demonstrate, this condition is necessary but insufficient.

Let us study the three-point bare coupling $\tilde{\phi}_{\mathbf{3}} \tilde{\phi}_{\mathbf{3}} \tilde{\phi}_{\mathbf{3}}$ in the $\mathbb{Z}_2$-gauged $A_4$ model. 
Since 
\begin{align}
\mathbf{3}\otimes\mathbf{3}\otimes\mathbf{3} \ni \mathbf{1},
\end{align}
the naive selection rule in Eq.~\eqref{eq:cond_orb} would suggest that a non-vanishing bare coupling is allowed after $\mathbb{Z}_2$-gauging. 
The allowed three-point couplings are explicitly written as
\begin{align}
	(\phi_{\mathbf{3}} \otimes \phi_{\mathbf{3}}\otimes \phi_{\mathbf{3}})_{\mathbf{1}} =
	\begin{cases}
		x_1(y_2 z_3 + z_2 y_3) + x_2(y_3 z_1 + z_3 y_1) + x_3(y_1 z_2 + z_1 y_2)\\
		x_1(y_2 z_3 - z_2 y_3) + x_2(y_3 z_1 - z_3 y_1) + x_3(y_1 z_2 - z_1 y_2)
	\end{cases},
	\label{eq:3*3*3}
\end{align}
where $x_i, y_i$, and $z_i$ denote the $i$-th components of the three triplet fields, respectively. 
Substituting Eq.~\eqref{eq:3_tilde} into Eq.~\eqref{eq:3*3*3}, we find that the corresponding three-point couplings of the projected triplets completely vanish:
\begin{align}
(\tilde\phi_{\mathbf{3}} \otimes \tilde\phi_{\mathbf{3}} \otimes \tilde\phi_{\mathbf{3}})_{\mathbf{1}} = 0,
\end{align}
because the first component of the $\mathbb{Z}_2$-invariant fields is identically zero. 
Thus, this bare coupling is eliminated by the $\mathbb{Z}_2$ projection and is absent from the tree-level effective Lagrangian.

This apparent contradiction can be understood more easily through the tensor decompositions of the $A_4$ group. 
According to the irreducible decomposition rules of $A_4$, the tensor product of two triplet representations is given by
\begin{align}
{\phi_\mathbf{3}} \otimes {\phi_\mathbf{3}} =  \phi_\mathbf{1} + \phi_{\mathbf{1}'} + \phi_{\mathbf{1}''} + \phi_\mathbf{3}^1 + \phi_\mathbf{3}^2,
\label{eq:3*3}
\end{align}
where two distinct triplet combinations $\phi_\mathbf{3}^{1}$ and $\phi_\mathbf{3}^{2}$ appear. Both belong to the same triplet representation $\mathbf{3}$ of $A_4$, but they possess different field components:
\begin{align}
\phi_{\mathbf 3}^1
=
\begin{pmatrix}
x_2 y_3 + x_3 y_2 \\
x_3 y_1 + x_1 y_3 \\
x_1 y_2 + x_2 y_1
\end{pmatrix}_{\mathbf 3}, 
\quad 
\phi_{\mathbf 3}^2
=
\begin{pmatrix}
x_2 y_3 - x_3 y_2 \\
x_3 y_1 - x_1 y_3 \\
x_1 y_2 - x_2 y_1
\end{pmatrix}_{\mathbf 3}.
\end{align}
Here, it immediately follows that the multiplication rules based on $A_4$ apparently contradict the transformation law under $H$ in Eq.~\eqref{eq:A4}. 
This is because $\phi_{\mathbf 3}^1$ transforms as $\phi_{\mathbf 3}^1 \to U_+\phi_{\mathbf 3}^1$, 
which is inconsistent with Eq.~\eqref{eq:A4}, while the other fields $\phi_{\mathbf 1}', \phi_{\mathbf 1}''$, 
and $\phi_{\mathbf 3}^2$ consistently transform according to Eq.~\eqref{eq:A4}.

From the perspective of $S_4 \cong A_4 \rtimes \mathbb{Z}_2$, it is evident that $\phi_{\mathbf 3}^1$ 
corresponds to the $\mathbf{3}$ representation of $S_4$, and the above multiplication rule for the triplets should actually be regarded as that of $S_4$:
\begin{align}
\mathbf{3}' \otimes\mathbf{3}' =  \mathbf{1} \oplus \mathbf{2} \oplus \mathbf{3} \oplus \mathbf{3}'.
\end{align}
This result clearly demonstrates that multiplication rules based solely on $G$-representations are generally not closed. 
Crucially, this inconsistency cannot be resolved simply by assigning an alternative field transformation such as $U=U_+$. 
Even if one were to construct a theory solely with fields transforming under $U_+$ (corresponding to the $\mathbf{3}$ of $S_4$), 
their tensor product would naturally generate fields transforming under $U_-$ (such as the $\mathbf{3}'$ of $S_4$). 
As the multiplication rule in Eq.~\eqref{eq:3*3} suggests, it is unavoidable that distinct fields belonging to the same irreducible representation of $G$ will transform differently under $H$. 
Therefore, a consistent description of these fields and their interactions necessitates extending our framework to the full semidirect product $G \rtimes H$.

This fundamental limitation reveals that selection rules based solely on $H$-orbits are insufficient for non-Abelian models. 
In the next section, we investigate how to obtain a selection rule that properly determines the allowed bare couplings under a general discrete $H$-gauging of an underlying non-Abelian global symmetry $G$.

\section{Discrete $H$-gauging for $G \rtimes H$}
\label{sec:rep}

We now investigate the $H$-gauging of models with $G \rtimes H$ symmetry, assuming both $G$ and $H$ are discrete groups. 
When $G=\mathbb{Z}_N$ and $H$ is a cyclic subgroup, the selection rule is governed by a fusion-like algebra based on the $H$-orbit multiplication of conjugacy classes of $G$. 
For general $G$ and $H$, however, there is no obvious one-to-one correspondence between irreducible representations and $H$-orbits of conjugacy classes.
In particular, when $G$ is non-Abelian, as shown in the previous section, 
distinct fields in the same irreducible representation $\mathbf{r}^{(\alpha)}$ of $G$ can transform differently under $H$. 
An $n$-point interaction $\phi_{\alpha_1} \phi_{\alpha_2} \cdots \phi_{\alpha_n}$ in the Lagrangian is projected onto the $H$-invariant sector as
\begin{align}
\phi_{\alpha_1} \phi_{\alpha_2} \cdots \phi_{\alpha_n} \to \tilde{\phi}_{\alpha_1} \tilde{\phi}_{\alpha_2} \cdots \tilde{\phi}_{\alpha_n}.
\end{align}
To determine which representations actually remain in the product of $H$-invariant fields, 
we analyze the tensor-product decomposition of the irreducible representations of $G\rtimes H$. 
We will extend this tensor decomposition to the $H$-invariant field components, allowing the character-based selection rules to be appropriately generalized to $H$-invariant fields. 
This approach yields a fusion-like rule corresponding to the irreducible decomposition of products of $H$-invariant fields. 
Consequently, we will demonstrate the relationship between these selection rules and fusion-like algebra restricted to $H$-invariant vectors.

In what follows, we assume a theory with a global $G\rtimes H$ symmetry to ensure consistency with $H$-gauging. 
We first review the coupling selection rule based on the $G\rtimes H$ symmetry, and subsequently extend this result to models with $H$-gauging.

\subsection{Selection Rule for Conventional Group Symmetry}

Let us first review the coupling selection rule for a conventional group symmetry $G\rtimes H$. 
Each field $\phi_\alpha$ is labeled by an irreducible representation $\mathbf{r}^{(\alpha)}$ of $G\rtimes H$. 
Since the theory is invariant under $G\rtimes H$, the tensor product $\mathbf{r}^{(\alpha_1)} \otimes \mathbf{r}^{(\alpha_2)} \otimes \cdots \otimes \mathbf{r}^{(\alpha_n)}$ is required to contain the trivial representation $\mathbf{1}$ in order for an $n$-point interaction $\phi_{\alpha_1} \phi_{\alpha_2} \cdots \phi_{\alpha_n}$ to have a non-vanishing coupling. 
Namely, the condition 
\begin{align}
\mathbf{r}^{(\alpha_1)} \otimes \mathbf{r}^{(\alpha_2)} \otimes \cdots \otimes \mathbf{r}^{(\alpha_n)} \ni \mathbf{1}
\end{align}
should be satisfied, which originates directly from the conventional group symmetry.
In fact, this selection rule can be reformulated purely based on group-theoretical arguments. 

Let us introduce the projection operator onto the irreducible representation $\mathbf{r}^{(\gamma)}$,
\begin{align}\label{eq:P1}
P^{(\gamma)} = \frac{\dim V^{(\gamma)}}{|G\rtimes H|} \sum_{g\in G\rtimes H} \chi^{(\gamma)*} (g) \rho(g),
\end{align}
where $V^{(\gamma)}$ is the representation space of $\mathbf{r}^{(\gamma)}$ and $\chi^{(\gamma)*}(g)$ is the complex conjugate of its character. 
From the great orthogonality theorem of finite groups, 
it is straightforward to see that $P^{(\gamma)}$ projects an arbitrary vector onto the subspace of the irreducible representation $\mathbf{r}^{(\gamma)}$.

We then consider the product of two representations and its decomposition into a direct sum of irreducible representations. 
Let $V^{(\alpha)}$ and $V^{(\beta)}$ be the representation spaces for $G\rtimes H$. 
Their tensor product decomposes as
\begin{align}
V^{(\alpha)} \otimes V^{(\beta)} = \bigoplus_{\gamma} N^{\gamma}_{\alpha \beta} V^{(\gamma)},
\label{eq:ab_in_g}
\end{align}
where $N^{\gamma}_{\alpha \beta}$ denotes the multiplicity of $\mathbf{r}^{(\gamma)}$ appearing in the product. 
Thus, for the tensor product representation of $\phi_{\alpha} \otimes \phi_{\beta}$, the corresponding projection operator 
is given by
\begin{align}\label{eq:P2}
P^{(\alpha \otimes \beta \to \gamma)} = \frac{\dim V^{(\gamma)}}{|G\rtimes H|} \sum_{g\in G\rtimes H} \chi^{(\gamma)*}(g) 
\left( \rho^{(\alpha)}(g) \otimes \rho^{(\beta)}(g) \right).
\end{align}
Taking the trace over $V^{(\alpha)} \otimes V^{(\beta)}$, we see that $\mathrm{Tr} [P^{(\alpha\otimes \beta \to \gamma)}]= \dim V^{(\gamma)} N^\gamma_{\alpha\beta}$ from Eq.~\eqref{eq:ab_in_g}. 
Evaluating the trace of Eq.~\eqref{eq:P2} then yields
\begin{align}\label{eq:Ngab}
N^\gamma_{\alpha\beta} = \frac{1}{|G\rtimes H|} 
\sum_{g\in G\rtimes H} \chi^{(\gamma)*} (g) \chi^{(\alpha)}(g) \chi^{(\beta)}(g).
\end{align}
This is a well-known character relation. 
This formula dictates the decomposition of tensor products, providing the group-theoretical foundation for evaluating matrix elements analogous to the Wigner-Eckart theorem. 
In particular, for the trivial singlet representation $\mathbf{r}^{(\gamma)} = \mathbf{1}$, since $\chi^{(\mathbf{1})}(g)=1$, the formula reduces to
\begin{align}
N^{\mathbf{1}}_{\alpha\beta} = \frac{1}{|G\rtimes H|} 
\sum_{g\in G\rtimes H} \chi^{(\alpha)}(g) \chi^{(\beta)}(g).
\end{align}
This counts the number of trivial singlet representations appearing in the tensor decomposition of $\phi_{\alpha} \otimes \phi_{\beta}$. 
Thus, a non-zero value, $N^{\mathbf{1}}_{\alpha\beta} \neq 0$, guarantees the existence of a non-vanishing two-point coupling in the singlet channel $(\phi_{\alpha} \otimes \phi_{\beta})_{\mathbf{1}}$.

The generalization to an $n$-point interaction is straightforward. 
We define the singlet projection operator for the tensor product $\phi_{\alpha_1} \otimes \phi_{\alpha_2} \otimes \cdots \otimes \phi_{\alpha_n}$ as 
\begin{align}\label{eq:P_singlet}
P^{(\mathbf{1})} = 
\frac{1}{|G\rtimes H|} \sum_{g\in G\rtimes H} \rho^{(\alpha_1)}(g) \otimes \rho^{(\alpha_2)}(g) \otimes \cdots \otimes \rho^{(\alpha_n)}(g).
\end{align}
By taking the trace of this operator, we obtain the necessary and sufficient condition for an allowed $n$-point interaction in a $G\rtimes H$ symmetric theory as $N^{\mathbf{1}}_{\alpha_1 \alpha_2 \cdots \alpha_n}  \neq 0$, that is 
\begin{align}\label{eq:selection_rule_GH}
\sum_{g\in G\rtimes H} \chi^{(\alpha_1)}(g) \chi^{(\alpha_2)}(g) \cdots \chi^{(\alpha_n)}(g) \neq 0.
\end{align}
This constitutes the selection rule for a conventional group symmetry $G\rtimes H$.

It is well known that this result can also be derived from the fusion algebra of the tensor product,
\begin{align}\label{eq:fusion}
\mathbf{r}^{(\alpha)} \otimes \mathbf{r}^{(\beta)} = \bigoplus_{\gamma} N^\gamma_{\alpha \beta} \mathbf{r}^{(\gamma)}.
\end{align}
Since the irreducible decomposition of the tensor product satisfies associativity, the multiplicity coefficients $N^\gamma_{\alpha \beta}$ also satisfy the associativity relation, 
\begin{align}
	\sum_{\sigma } N^\sigma_{\alpha \beta} N^\delta_{\sigma \gamma} = 
	\sum_{\sigma } N^\delta_{\alpha \sigma} N^\sigma_{\beta \gamma} = N^{\delta}_{\alpha \beta \gamma}.
\end{align}
Using this relation, we can recursively obtain the selection rule in Eq.~\eqref{eq:selection_rule_GH}.

\subsection{Selection Rule for $H$-gauged Theories}

We now consider discrete $H$-gauging. 
After $H$-gauging, physical states are restricted to $H$-invariant vectors. 
We use the same notation $\tilde{\phi}_\alpha$ to denote the $H$-invariant fields as in the $G$-symmetric models. 
Since fields are labeled by representations of $G \rtimes H$, $\phi_\alpha$ transforms as $\rho^{(\alpha)}(h) \phi_\alpha$ under $H$. 
Thus, the $H$-invariant field in Eq.~\eqref{eq:H-invariant} should be modified as 
\begin{align}
\tilde{\phi}_\alpha =& P_H^{(\alpha)} \phi_\alpha,
\label{eq:tilde_phi_def}
\end{align}
where the projection operator in Eq.~\eqref{eq:P} is given by 
\begin{align}\label{eq:P_H}
P_H^{(\alpha)} = \frac{1}{|H|} \sum_{h \in H}\rho^{(\alpha)}(h).
\end{align}
Because physical states are restricted to the $H$-invariant states $\tilde{\phi}_\alpha$, $P_H^{(\alpha)}$ projects the representation space $V^{(\alpha)}$ onto the subspace spanned by $H$-invariant vectors. 
The representation space $V^{(\alpha)}$ thus decomposes as 
\begin{align}
V^{(\alpha)} &= \tilde{V}^{(\alpha)} \oplus \tilde V_\perp^{(\alpha)},
\end{align}
where $\tilde{V}^{(\alpha)}=P_H^{(\alpha)} V^{(\alpha)}$ is the subspace spanned by $H$-invariant vectors, and $\tilde V_\perp^{(\alpha)}=(1 - P_H^{(\alpha)}) V^{(\alpha)}$ denotes its orthogonal complement. 
Hereafter, we simply refer to $\tilde{V}^{(\alpha)}$ as the $H$-invariant subspace.
We also introduce projected characters defined as
\begin{align}\label{eq:chi-projected}
\tilde{\chi}^{(\alpha)}(g) = \mathrm{Tr}[\rho^{(\alpha)}(g) P_H^{(\alpha)}], 
\quad \quad 
\tilde{\chi}_\perp^{(\alpha)}(g) = \mathrm{Tr}[\rho^{(\alpha)}(g) (1-P_H^{(\alpha)})]. 
\end{align}
It can be shown that these projected characters also satisfy the following orthogonality relations\footnote{See Appendix \ref{sec:tilde_rho}.}: 
\begin{align}
\frac{1}{|G\rtimes H|} \sum_{g\in G \rtimes H} \tilde{\chi}^{(\alpha)*}(g) \tilde{\chi}^{(\beta)}(g) &=  
\frac{\mathrm{dim}\,\tilde{V}^{(\alpha)}}{\mathrm{dim}\,V^{(\alpha)}} \delta_{\alpha \beta},
\quad
\sum_{g\in G\rtimes H} \tilde{\chi}^{(\alpha)*}(g) \tilde{\chi}_\perp^{(\beta)}(g) =  0.
\label{eq:orthogonality}
\end{align}

The singlet-projection procedure for $G\rtimes H$ can be adapted to the $H$-invariant vectors. 
Similar to Eq.~\eqref{eq:P1}, to extract $\tilde{V}^{(\gamma)}$ components from $\tilde{V}^{(\alpha)}$, 
we propose a projection-like operator acting on an $H$-invariant field $\tilde{\phi}_\alpha$ as
\begin{align}
\tilde{P}^{(\alpha\to\gamma)} 
= \frac{\dim V^{(\gamma)}}{|G\rtimes H|} \sum_{g\in G\rtimes H} 
\tilde{\chi}^{(\gamma)*} (g) \rho^{(\alpha)}(g) P_H^{(\alpha)}.
\end{align}
Using the orthogonality relation for the projected characters in Eq.~\eqref{eq:orthogonality} and taking the trace of both sides, 
we obtain 
\begin{align}
\mathrm{Tr} [\tilde{P}^{(\alpha\to\gamma)}] 
&= \frac{\dim V^{(\gamma)}}{|G\rtimes H|} \sum_{g\in G\rtimes H} 
\tilde{\chi}^{(\gamma)*} (g) \tilde{\chi}^{(\alpha)}(g) 
\notag \\ 
&= \mathrm{dim}\,\tilde{V}^{(\gamma)} \delta_{\alpha\gamma}. \label{eq:trP}
\end{align}
It is important to note that $\mathrm{Tr} [\tilde{P}^{(\alpha\to\gamma)}]$ is exactly equal to $\mathrm{dim}\,\tilde{V}^{(\gamma)}$ (for $\alpha=\gamma$), which correctly counts the remaining vector components in $\mathbf{r}^{(\gamma)}$ after $H$-gauging. 
This property fully satisfies our requirement for establishing a necessary and sufficient condition for the selection rules.

For a tensor product of two $H$-invariant fields $\tilde{\phi}_{\alpha} \otimes \tilde{\phi}_{\beta}$, a projection-like operator $\tilde{P}^{(\alpha \otimes \beta \to \gamma)}$ that extracts the $\tilde{V}^{(\gamma)}$ component from $\tilde{V}^{(\alpha)}\otimes \tilde{V}^{(\beta)}$ is given by
\begin{align}\label{eq:Ptilde}
\tilde{P}^{(\alpha \otimes \beta \to \gamma)} 
= \frac{\dim V^{(\gamma)}}{|G\rtimes H|} \sum_{g\in G\rtimes H} 
\tilde{\chi}^{(\gamma)*}(g)
\left( \rho^{(\alpha)}(g) P_H^{(\alpha)} \otimes \rho^{(\beta)}(g) P_H^{(\beta)} \right).
\end{align}
Using the property $\mathrm{Tr}[A\otimes B]= \mathrm{Tr} [A]\cdot \mathrm{Tr} [B]$ and taking the trace of both sides, we obtain 
\begin{align}
\mathrm{Tr} [\tilde{P}^{(\alpha \otimes \beta \to \gamma)}] 
= 
\dim V^{(\gamma)} 
\tilde{N}^\gamma_{\alpha\beta},
\end{align}
where the parameter $\tilde{N}^\gamma_{\alpha\beta}$ is defined analogously to Eq.~\eqref{eq:Ngab} as 
\begin{align}\label{eq:tildeNgab}
\tilde{N}^\gamma_{\alpha\beta} = \frac{1}{|G\rtimes H|} \sum_{g\in G\rtimes H} 
\tilde{\chi}^{(\gamma)*}(g)
\tilde{\chi}^{(\alpha)}(g) \tilde{\chi}^{(\beta)}(g).
\end{align}
For the trivial singlet representation $\mathbf{r}^{(\gamma)}= \mathbf{1}$, since $\tilde{\chi}^{(\mathbf{1})}(g)=1$, we obtain 
\begin{align}\label{eq:tildeNgab_singlet}
\tilde{N}^{\mathbf{1}}_{\alpha\beta}  = \mathrm{Tr}[\tilde{P}^{(\alpha \otimes \beta \to \mathbf{1})}] 
= \frac{1}{|G\rtimes H|} \sum_{g\in G\rtimes H} 
\tilde{\chi}^{(\alpha)}(g) \tilde{\chi}^{(\beta)}(g).
\end{align}

The above relations can be straightforwardly generalized. 
For a tensor product involving an $n$-point interaction $\tilde{\phi}_{\alpha_1} \otimes \tilde{\phi}_{\alpha_2} \otimes \cdots \otimes \tilde{\phi}_{\alpha_n}$, we define 
\begin{align}\label{eq:tildeNgan}
\tilde{N}^\gamma_{\alpha_1\alpha_2\cdots\alpha_n} = 
\frac{1}{|G\rtimes H|} \sum_{g\in G\rtimes H} \tilde{\chi}^{(\gamma)*}(g) \tilde{\chi}^{(\alpha_1)}(g) \tilde{\chi}^{(\alpha_2)}(g) \cdots \tilde{\chi}^{(\alpha_n)}(g).
\end{align}
A projection-like operator that extracts the trivial singlet component is then given by 
\begin{align}\label{eq:tildeP_singlet}
\tilde{P}^{(\mathbf{1})} = \frac{1}{|G\rtimes H|} \sum_{g\in G\rtimes H} 
\rho^{(\alpha_1)}(g) P_H^{(\alpha_1)} \otimes
\rho^{(\alpha_2)}(g) P_H^{(\alpha_2)} 
\otimes \cdots \otimes
\rho^{(\alpha_n)}(g) P_H^{(\alpha_n)}. 
\end{align}
From both equations, we see $\mathrm{Tr}[\tilde{P}^{(\mathbf{1})}] = \tilde{N}^{\mathbf{1}}_{\alpha_1\alpha_2\cdots\alpha_n}$.

It is worth noting that the quantity $\tilde{N}^{\mathbf{1}}_{\alpha_1\alpha_2\cdots\alpha_n}$ is generally not an integer due to the multiple insertions of the projection operator $P_H$ (which is why we refer to $\tilde{P}$ as a projection-like operator). 
For instance, in the case of a two-field tensor product, this non-integer value implies that the entire space of $\tilde{V}^{(\gamma)}$ cannot be spanned solely by the $H$-invariant vectors in $\tilde{V}^{(\alpha)} \otimes \tilde{V}^{(\beta)}$, unlike the case of $\tilde{P}^{(\alpha\to\gamma)}$ in Eq.~\eqref{eq:trP}\footnote{Geometrically, $\tilde{N}^{\gamma}_{\alpha \beta}$ corresponds to the sum of the squared cosines of the principal angles between the corresponding subspaces \cite{Miao:1992va, 1361981468336614144}.}. 
Nevertheless, $\tilde{N}^{\mathbf{1}}_{\alpha_1\alpha_2\cdots\alpha_n} \neq 0$ guarantees the existence of a non-vanishing $n$-point interaction in the singlet channel of $\tilde{\phi}_{\alpha_1} \otimes \tilde{\phi}_{\alpha_2} \otimes \cdots \otimes \tilde{\phi}_{\alpha_n}$, since this implies there exists at least one vector component in the tensor product of $H$-invariant fields that 
has a nonzero projection onto the trivial-singlet subspace.
In fact, the quantity $\tilde{N}^\gamma_{\alpha_1\alpha_2\cdots\alpha_n}$ can be related to the Clebsch-Gordan (CG) coefficients for the $H$-invariant fields, as will be analyzed in detail in the following section.

Consequently, we derive the necessary and sufficient condition for a non-vanishing $n$-point coupling $\tilde{\phi}_{\alpha_1} \tilde{\phi}_{\alpha_2} \cdots \tilde{\phi}_{\alpha_n}$, which is allowed if and only if 
\begin{align}\label{eq:selection_rule_gauging}
\sum_{g\in G\rtimes H} \tilde{\chi}^{(\alpha_1)}(g) \tilde{\chi}^{(\alpha_2)}(g) \cdots \tilde{\chi}^{(\alpha_n)}(g)  \neq 0.
\end{align}
This result represents one of the main highlights of this work, establishing a rigorous non-invertible selection rule for $H$-gauged models with an underlying $G\rtimes H$ symmetry. 
Notably, this formula is completely general and does not rely on any Abelian assumptions; it is entirely valid for arbitrary discrete groups $G\rtimes H$, 
including cases with generally non-Abelian $H$-gauging.

\subsection{Fusion-like Algebra for $H$-gauged Theories}
\label{sec:Fusion-like Algebra}

Here, we discuss a possible construction of a fusion-like algebra for models with $H$-gauging that correctly reproduces the tree-level coupling selection rule in Eq.~\eqref{eq:selection_rule_gauging}, and we clarify the relationship between this fusion-like algebra and the selection rule. 
Unlike the fusion rule in Eq.~\eqref{eq:fusion} used for conventional group symmetries, the analogous decomposition of $H$-invariant subspaces,
\begin{align}
\tilde{V}^{(\alpha)} \otimes \tilde{V}^{(\beta)} = \bigoplus_{\gamma} \tilde{N}^{\gamma}_{\alpha\beta} \tilde{V}^{(\gamma)},
\end{align}
does not constitute a well-defined fusion-like algebra. 
This is because the coefficients $\tilde{N}^{\gamma}_{\alpha\beta}$ generally do not satisfy the associativity condition due to the multiple insertions of the projection operator $P_H$. 
As stated in the previous section, $\tilde{N}^{\gamma}_{\alpha \beta} \neq 0$ simply implies a non-zero overlap between the target space $\tilde{V}^{(\gamma)}$ and the tensor product $\tilde{V}^{(\alpha)} \otimes \tilde{V}^{(\beta)}$, rather than meaning the space is fully spanned.

Furthermore, after $H$-gauging, the original group symmetries $G\rtimes H$ and $G$ are generally broken, meaning that the irreducible representation $\mathbf{r}^{(\alpha)}$ is also decomposed. 
Consequently, it is evident that a fusion-like algebra should be defined for each individual vector component within the $H$-invariant subspace of $\mathbf{r}^{(\alpha)}$, rather than for the multiplet $\tilde{\phi}_\alpha$ as a whole. 
Such a component-wise fusion-like algebra is particularly useful for constructing low-energy effective theories that incorporate quantum effects, where tree-level coupling selection rules can be modified by quantum corrections~\cite{Kaidi:2024wio}. 
Even at tree level, as we will show, this detailed component analysis reveals characteristic features among the multiplet interactions.

Let $\{ \ket{\alpha_i} \}_{i= 1, \cdots , \dim V^{(\alpha)}}$ be an orthonormal basis of $V^{(\alpha)}$\footnote{Note that $\ket{\alpha_i}$ need not be a quantum state nor an eigenstate of a specific physical operator; it simply denotes a basis vector of the representation space.}. 
Using this basis, $\phi_\alpha$ can be expanded as $\phi_\alpha = \ket{\alpha_i} (\phi_\alpha)_i$, where $(\phi_\alpha)_i$ is the $i$-th component of the field $\phi_\alpha$, and summation over repeated indices is implied. 
The $H$-invariant field $\tilde{\phi}_\alpha$ is then written as $\tilde{\phi}_\alpha = P_H^{(\alpha)} \phi_\alpha = \ket{\alpha_i} P^{(\alpha)}_{H, ij} (\phi_\alpha)_j$, where $P^{(\alpha)}_{H, ij}$ is the matrix representation of the projection operator given in Eq.~\eqref{eq:P_H}. 
We define the $H$-invariant basis $\ket{\tilde{\alpha}_i}$ and its corresponding components $(\tilde{\phi}_\alpha)_i$ as
\begin{align}
	\ket{\tilde{\alpha}_i} 
	\equiv \ket{\alpha_j} P^{(\alpha)}_{H, ji}, 
	\quad
	(\tilde{\phi}_\alpha)_i 
	\equiv  P^{(\alpha)}_{H, ij} (\phi_\alpha)_j,
\end{align}
so that the $H$-invariant field is expressed as $\tilde{\phi}_\alpha = \ket{\tilde{\alpha}_i} (\tilde{\phi}_\alpha)_i $. 
The inner product of $\ket{\tilde{\alpha}_i}$ and $\ket{\tilde{\alpha}_j}$ is given by
\begin{align}
	\bra{\tilde{\alpha}_i} \tilde{\alpha}_j \rangle = P_{H, ij}^{(\alpha)}. 
\end{align} 
Thus, they become strictly orthogonal to each other when $P_H^{(\alpha)}$ is diagonalized. 
We hereafter work in a basis where $P_H^{(\alpha)}$ is diagonal.

Tensor products of $H$-invariant vectors should be $H$-invariant. Hence, the product of two $H$-invariant fields decomposes as 
\begin{align}
\tilde{\phi}_\alpha \otimes \tilde{\phi}_\beta = 
\ket{\tilde{\alpha}_i} \otimes \ket{\tilde{\beta}_j} (\tilde{\phi}_\alpha)_i  (\tilde{\phi}_\beta)_j = 
\sum_{\gamma} \sum_{a = 1}^{N^\gamma_{\alpha \beta}} \ket{\tilde{\gamma}_k^{a}} (\tilde{\phi}_\gamma)_k^a,
\end{align}
where we account for the multiplicity of representations. The index $a$ distinguishes these multiple copies, ensuring the uniqueness of the expansion coefficients $(\tilde{\phi}_\gamma)_k^a$. 
By extracting the components, we obtain the multiplication rule derived from the tensor decomposition of $G\rtimes H$, given by
\begin{align}\label{eq:expansion}
(\tilde{\phi}_\alpha)_i (\tilde{\phi}_\beta)_j =
\sum_{\gamma} \sum_{a = 1}^{N^\gamma_{\alpha \beta}} \bra{\tilde{\alpha}_i \tilde{\beta}_j} \ket{\tilde{\gamma}_k^a} 
(\tilde{\phi}_\gamma)_k^a,
\end{align}
where the inner product $\bra{\tilde{\alpha}_i \tilde{\beta}_j}\ket{\tilde{\gamma}_k^a}$ corresponds to the CG coefficient for the $H$-invariant vector spaces. 
Because the underlying tensor decomposition is associative, the CG coefficients restricted to $H$-invariant vectors inherit this associativity. 
As noted earlier, the vector spaces $\tilde{V}^{(\alpha)}$ themselves do not form a closed multiplication algebra. Instead, we emphasize that it is precisely this component-wise expansion in Eq.~\eqref{eq:expansion} that constitutes the defining fusion-like algebra for our $H$-gauged theory, providing a rigorously defined and associative algebraic structure.

As usual, the CG coefficients contain sufficient information to determine the coupling selection rules. 
For instance, $\bra{\tilde{\alpha}_i \tilde{\beta}_j} \ket{\tilde{\gamma}_k^a} \neq 0$ for at least one $a$ implies that there exists a non-zero component of $\tilde{V}^{(\gamma)}$ within the tensor product of the two vectors $\ket{\tilde{\alpha}_i}$ and $\ket{\tilde{\beta}_j}$, even though, in general, $\tilde{V}^{(\gamma)} \not\subset \tilde{V}^{(\alpha)} \otimes \tilde{V}^{(\beta)}$. 
Furthermore, the projection-like operator in Eq.~\eqref{eq:Ptilde} can be expressed using these basis vectors as
\begin{align}
\tilde{P}^{(\alpha \otimes \beta \to \gamma)} 
&= 
P^{(\alpha \otimes \beta \to \gamma)}
\circ \left( P_H^{(\alpha)} \otimes P_H^{(\beta)} \right)
\notag
\\
&= 
\sum_{i, j, k}
\ket{\gamma_i} \bra{\gamma_i} \cdot 
\left( \ket{\tilde{\alpha}_j} \bra{\tilde{\alpha}_j} \otimes \ket{\tilde{\beta}_k} \bra{\tilde{\beta}_k} \right).
\label{eq:ab->c}
\end{align}
Taking the trace of both sides, we find
\begin{align}
\tilde{N}^{\gamma}_{\alpha \beta} = \frac{1}{\dim V^{(\gamma)}} 	
\sum_{i, j, k}
\left|\bra{\tilde{\alpha}_j \tilde{\beta}_k}\ket{\tilde{\gamma}_i}\right|^2.
\label{eq:N_CG_relation}
\end{align}
Thus, we see that $\tilde{N}^{\gamma}_{\alpha \beta} \geq 0$ is a positive semi-definite and basis-independent quantity, given by the sum of the absolute squares of the CG coefficients over the orthogonal basis, whereas the individual $H$-invariant CG coefficients are basis-dependent. 
Consequently, as stated previously, a non-zero $\tilde{N}^{\gamma}_{\alpha \beta}$ implies a physical overlap between $\tilde{V}^{(\gamma)}$ and $\tilde{V}^{(\alpha)} \otimes \tilde{V}^{(\beta)}$, serving as the criterion for allowed interactions.

The above results can be straightforwardly extended to $n$-point interactions. 
The parameter $\tilde{N}^{\gamma}_{\alpha_1 \alpha_2 \cdots \alpha_n}$ defined in Eq.~\eqref{eq:tildeNgan} can be expressed via these inner products as  
\begin{align}
\tilde{N}^{\gamma}_{\alpha_1 \alpha_2 \cdots \alpha_n} =  \frac{1}{\dim V^{(\gamma)}} 
\sum_{\{i\}, l, a} \left|\bra{(\tilde{\alpha}_1)_{i_1} (\tilde{\alpha}_2)_{i_2} \cdots (\tilde{\alpha}_n)_{i_n}} \ket{\tilde{\gamma}_l^a} \right|^2.
\label{eq:tilde_N^g_a...a}
\end{align}
A component of an $n$-point interaction $(\tilde{\phi}_{\alpha_1})_{i_1} (\tilde{\phi}_{\alpha_2})_{i_2} \cdots (\tilde{\phi}_{\alpha_n})_{i_n}$ is non-vanishing if and only if  
$\bra{(\tilde{\alpha}_1)_{i_1} (\tilde{\alpha}_2)_{i_2} \cdots (\tilde{\alpha}_n)_{i_n}} \ket{\mathbf{1}} \neq 0$.
Therefore, the full $n$-point interaction $\tilde{\phi}_{\alpha_1} \otimes \tilde{\phi}_{\alpha_2} \otimes \cdots \otimes \tilde{\phi}_{\alpha_n}$ possesses a non-vanishing coupling if there exists at least one combination of components such that
\begin{align}\label{eq:selection_rule_f}
\bra{(\tilde{\alpha}_1)_{s_1} (\tilde{\alpha}_2)_{s_2} \cdots (\tilde{\alpha}_n)_{s_n}} \ket{\mathbf{1}} \neq 0,
\quad s_j=1,\cdots, \dim \tilde V^{(\alpha_j)}.
\end{align}
This constitutes a sufficient condition for non-vanishing couplings after $H$-gauging, 
generalizing to the non-Abelian case while smoothly incorporating the selection rules 
of Eq.~\eqref{eq:selection_rule} for Abelian $\mathbb{Z}_N$ group symmetries. 
It is also evident that summing the squared CG coefficients over all indices precisely 
reproduces the character-based selection rule derived in Eq.~\eqref{eq:selection_rule_gauging}.

Let us briefly comment on the relationship between our results and models with an Abelian $G$ commonly studied in the literature. 
If all $H$-invariant spaces are one-dimensional, $\tilde{N}^{\gamma}_{\alpha \beta} \neq 0$ strictly dictates that $\tilde{V}^{(\gamma)}$ can be fully spanned by the $H$-invariant vectors $\tilde{V}^{(\alpha)} \otimes \tilde{V}^{(\beta)}$; namely, $\tilde{V}^{(\gamma)} \subset P_H^{(\gamma)} (\tilde{V}^{(\alpha)} \otimes \tilde{V}^{(\beta)})$ due to the dimensionality constraints. 
In such cases, a consistent fusion-like algebra can be directly constructed via $\tilde{N}^{\gamma}_{\alpha \beta}$. 
This situation actually occurs when $G$ is Abelian. Here, the associativity of the irreducible decomposition of the tensor product is satisfied in the sense that
\begin{align}
\tilde{\phi}_{\alpha} \otimes \tilde{\phi}_{\beta} \simeq \sum_{\gamma \in \{\gamma \mid \tilde{N}^{\gamma}_{\alpha \beta} \neq0\}} 
\tilde{\phi}_{\gamma},
\end{align}
where the right-hand side is understood as a formal sum over all allowed irreducible representations up to their respective coefficients. 
In these models, the selection rules for $n$-point couplings can be deduced solely from $\tilde{N}^{\alpha}_{\beta \gamma}$, as previously demonstrated for the $D_N$ model.

%%%%%%%%%%%%%%%%%%%%
%%%%%%%%%%%%%%%%%%%%
%%%%%%%%%%%%%%%%%%%%

\section{Double Cosets for Projected Characters and Examples}
\label{sec:exmple}

In this section, we present several concrete examples of models with the discrete $H$-gauging of an underlying $G\rtimes H$ group symmetry. 

To obtain the non-invertible selection rule in Eq.~\eqref{eq:selection_rule_gauging}, 
we explicitly evaluate the projected characters $\tilde{\chi}^{(\alpha)}$ for all the irreducible representations $\mathbf{r}^{(\alpha)}$ of $G\rtimes H$. 
We note that, although the projected characters $\tilde{\chi}^{(\alpha)}$ no longer respect the conjugacy classes of $G \rtimes H$, 
distinct projected character values are instead classified by the double coset decomposition of $G\rtimes H$ with respect to $H$, given by
\begin{align}
	G \rtimes H =  \bigsqcup_i H g_i H,
\end{align}
where $g_i$ denotes a representative element of each double coset. 
The projected character $\tilde{\chi}^{(\alpha)}(g)$ takes the same value for any $g$ and $g'$ contained in the same double coset $H g_i H$. 
This is because if $g'$ and $g$ belong to the same double coset, $g'$ can be expressed as $g' = h_1 g h_2$ for some $h_1, h_2 \in H$, and we find
\begin{align}
	\tilde{\chi}^{(\alpha)}(g') &= \mathrm{Tr} \left[ \rho^{(\alpha)}(g') P_H^{(\alpha)} \right] \notag \\
	&= \mathrm{Tr} \left[ \rho^{(\alpha)}(h_1) \rho^{(\alpha)}(g) \rho^{(\alpha)}(h_2) P_H^{(\alpha)} \right] \notag \\
	&= \mathrm{Tr} \left[ \rho^{(\alpha)}(g) P_H^{(\alpha)} \right] \notag \\
	&= \tilde{\chi}^{(\alpha)}(g).
\end{align}
Here, we used the fact that $\rho^{(\alpha)}(h) P_H^{(\alpha)} = P_H^{(\alpha)} \rho^{(\alpha)}(h) = P_H^{(\alpha)}$. 
Consequently, the multiplicity parameter $\tilde{N}^{\gamma}_{\alpha_1 \alpha_2 \cdots \alpha_n}$ can be efficiently calculated as
\begin{align}\label{eq:tildeNgan_double_coset}
\tilde{N}^\gamma_{\alpha_1\alpha_2\cdots\alpha_n} = 
	\frac{1}{|G\rtimes H|} \sum_{g_i} |H g_i H| \tilde{\chi}^{(\gamma)*}(g_i) \tilde{\chi}^{(\alpha_1)}(g_i) \tilde{\chi}^{(\alpha_2)}(g_i) \cdots \tilde{\chi}^{(\alpha_n)}(g_i),
\end{align}
where $|H g_i H|$ is the number of elements contained in the double coset $H g_i H$.

We also show the fusion-like algebra defined in Eq.~\eqref{eq:expansion}, namely the multiplication rules between $H$-invariant fields based on their component-wise expansions. 
By using this fusion-like algebra recursively, we can alternatively derive the non-invertible selection rules. 
The results obtained from these two distinct formulas are consistent, as expected.

\subsection{$\mathbb{Z}_2$-gauging of $D_N \cong \mathbb{Z}_N \rtimes \mathbb{Z}_2$}

Let us study the non-invertible selection rules for the $\mathbb{Z}_2$-gauging of the $D_N$ symmetry group. 
Since $D_N$ is isomorphic to $\mathbb{Z}_N \rtimes \mathbb{Z}_2$, 
its non-invertible selection rules should be equivalent to those studied in Section \ref{sec:gauging_of_Z_N}. 
Here, we reconsider this equivalence from the perspective of the $\mathbb{Z}_2$-invariant representations of $D_N$ 
as an illustrative example.

For even $N$, the group $D_N$ possesses doublets $\mathbf{2}_q$ with $q = 1, \dots, N/2 - 1$ and four singlets $\mathbf{1}_{++}, \mathbf{1}_{+-}, \mathbf{1}_{-+}, \mathbf{1}_{--}$. 
For odd $N$, it has doublets $\mathbf{2}_q$ with $q = 1, \dots, (N - 1)/2$ and two singlets $\mathbf{1}_+, \mathbf{1}_{-}$. 
Let $h$ denote the generator of $\mathbb{Z}_2$. The matrix representations of $h$ for the singlets are given by
\begin{align}
	\rho^{(\mathbf{1}_{++})}(h) &= \rho^{(\mathbf{1}_{+-})}(h) = 1, &
	\rho^{(\mathbf{1}_{-+})}(h) &= \rho^{(\mathbf{1}_{--})}(h) = -1,
	\notag \\
	\rho^{(\mathbf{1}_{+})}(h)  &= 1, &
	\rho^{(\mathbf{1}_{-})}(h) &= -1,	
\end{align}
and for the doublets, in both the even and odd $N$ cases, the representation is
\begin{align}
	\rho^{(\mathbf{2}_q)}(h) = 
	\begin{pmatrix}
	0 & 1 \\
	1 & 0
	\end{pmatrix}.
\end{align}
After $\mathbb{Z}_2$-gauging, the representations with negative parity, namely $\mathbf{1}_{-+}, \mathbf{1}_{--}$, and $\mathbf{1}_{-}$, are projected out.
Since the field multiplet transforming in the doublet representation $\mathbf{2}_{q}$ of $D_N$ is written as 
\begin{align}
\Phi_{\mathbf{2}_{q}} = 
\begin{pmatrix}
\phi_q \\ \phi_{-q}
\end{pmatrix},
\end{align}
applying the $H$-invariant projection defined in Eq.~\eqref{eq:tilde_phi_def} yields
\begin{align}
\tilde{\Phi}_{\mathbf{2}_{q}} = \frac{1}{2}
\begin{pmatrix}
\phi_q + \phi_{-q} \\ \phi_q + \phi_{-q}
\end{pmatrix} = \tilde{\phi}_{\mathbf{2}_q} \ket{\tilde{\mathbf{2}}_q},
\end{align}
where $\tilde{\phi}_{\mathbf{2}_q}$ on the right-hand side denotes the scalar field component, and $\ket{\tilde{\mathbf{2}}_q}$ denotes its normalized basis vector. 
To ensure the field is canonically normalized, $\tilde{\phi}_{\mathbf{2}_q}$ and $\ket{\tilde{\mathbf{2}}_q}$ are respectively defined as 
\begin{align}
	\tilde{\phi}_{\mathbf{2}_q} = \frac{1}{\sqrt{2}} (\phi_{q} + \phi_{-q}),
	\quad \quad
	\ket{\tilde{\mathbf{2}}_q}  = 
	\begin{pmatrix}
		\frac{1}{\sqrt{2}} \\
		\frac{1}{\sqrt{2}}
	\end{pmatrix}.
\end{align}

Projected characters $\tilde \chi^{(\alpha)}$ for $D_N$ representations are double coset functions.
Since $\mathbb{Z}_N$ is Abelian, the double coset decomposition of $D_N$ with respect to $\mathbb{Z}_2$ is given by $\mathbb{Z}_2$-orbit sum of each conjugacy class restricted to $\mathbb{Z}_N$ elements.
They are given by
\begin{align}
	\tilde C_1 &= \{e, h  \} = C_1 \cup C_1h,
	\notag\\
	\tilde C_2^{(1)} &= \{a, a^{N-1}, ah , a^{N-1}h  \} = C_2^{(1)} \cup C_2^{(1)}b,
	\notag\\
	&\vdots 
	\notag\\
	\tilde C_2^{(m)} &= \{a^m, a^{N-m}, a^mh, a^{N-m}h \} = C_2^{(m)} \cup C_2^{(m)}b,
	\notag\\
	&\vdots 
	\notag\\
	\tilde C_2^{(\frac N 2 -1)} &= \{a^{\frac N 2 -1}, a^{\frac N 2 +1}, a^{\frac N 2 -1}h, a^{\frac N 2 +1}h \} = C_2^{(\frac N 2 -1)} \cup C_2^{(\frac N 2 -1)}h,
	\notag\\
	\tilde C_1' &= \{a^{\frac N2}, a^{\frac N2} h \} = C_1' \cup C_1' h,
\end{align}
for even $N$ and
\begin{align}
	\tilde C_1 &= \{e, h  \} = C_1 \cup C_1h,
	\notag\\
	\tilde C_2^{(1)} &= \{a, a^{N-1}, ah , a^{N-1}h  \} = C_2^{(1)} \cup C_2^{(1)}h,
	\notag\\
	&\vdots 
	\notag\\
	\tilde C_2^{(m)} &= \{a^m, a^{N-m}, a^mh, a^{N-m}h \} = C_2^{(m)} \cup C_2^{(m)}h,
	\notag\\
	&\vdots 
	\notag\\
	\tilde C_2^{(\frac{N-1}2)} &= \{a^{\frac{N-1}2}, a^{\frac{N+1}2}, a^{\frac{N-1}2}h, a^{\frac{N+1}2}h \} = C_2^{(\frac N 2 -1)} \cup C_2^{(\frac N 2 -1)}h,
\end{align}
for odd $N$.
$C_2^{(m)}, C_1$ and $C_1'$ denotes the usual conjugacy classes of $D_N$.
There are $N/2 + 1$ double cosets for even $N$, and $(N+1)/2$ double cosets for odd $N$.
Thus the number of double coset corresponds to the number of remaining representations after gauging.
The projected characters for even and odd $N$ are summarized in Tables \ref{tab:Chara_even_DN} and \ref{tab:Chara_odd_DN}, respectively. 
\begin{table}[htb!]
\begin{center}
\renewcommand{\arraystretch}{1.5} 
\setlength{\tabcolsep}{5pt} 
\begin{tabular}{| c |   c  c  c |}
\hline
 & $\tilde{\chi}^{(\mathbf{1}_{++})}$ & $\tilde{\chi}^{(\mathbf{1}_{+-})}$ & $\tilde{\chi}^{(\mathbf{2}_{q})}$\\ 
\hline
$\tilde{C}_1$ & $1$& $1$ & $1$  
\\
$\tilde C_2^{(1)}$ & $1$ & $-1$ & $\cos \frac{2\pi q}{N}$
\\
$\vdots$ & $\vdots$ & $\vdots$ & $\vdots$
\\
$\tilde C_2^{(m)}$& $1$ & $(-1)^m$ & $\cos \frac{2\pi qm}{N}$ 
\\
$\vdots$  & $\vdots$ & $\vdots$ & $\vdots$
\\
$\tilde{C}_2^{(\frac N2 -1)}$ & $1$ & $(-1)^{\frac N2 -1}$ & $\cos \frac{2\pi q}{N}(\frac{N}2 -1)$
\\
$\tilde{C}_1'$ & $1$ & $(-1)^{\frac N2}$ & $(-1)^q$
\\
\hline
\end{tabular}
\end{center}
\vspace{-5mm} \caption{
$\mathbb{Z}_2$ projected characters of $D_N$ representations with even $N$.
$q$ runs from $1$ to $N/2 -1$.
$\tilde C_1,  \tilde C_2^{(m)}, \tilde C_1'$ denote the $\mathbb{Z}_2$-orbit sum of the conjugacy classes of $D_N$, i.e., $\tilde C_1 = C_1 \cup C_1 h, \tilde C_2^{(m)} = C_2^{(m)} \cup C_2^{(m)}h, \tilde C_1' =C_1' \cup C_1' h$, and hence $|\tilde C_1| =  |\tilde C_1'| = 2$, and $|\tilde C_2^{(m)}| = 4$.}
\label{tab:Chara_even_DN}
\end{table}
\begin{table}[htb!]
\begin{center}
\renewcommand{\arraystretch}{1.5} 
\setlength{\tabcolsep}{5pt} 
\begin{tabular}{| c |   c   c |}
\hline
 & $\tilde{\chi}^{(\mathbf{1}_{+})}$ & $\tilde{\chi}^{(\mathbf{2}_{q})}$\\ 
\hline
$\tilde{C}_1$ & $1$&  $1$  
\\
$\tilde C_2^{(1)}$ & $1$ &  $\cos \frac{2\pi q}{N}$
\\
$\vdots$ & $\vdots$ &  $\vdots$
\\
$\tilde C_2^{(m)}$& $1$ &  $\cos \frac{2\pi qm}{N}$ 
\\
$\vdots$  & $\vdots$ &  $\vdots$
\\
$\tilde{C}_2^{(\frac{N -1}2)}$ & $1$ &  $\cos \frac{2\pi q}{N}(\frac{N -1}2)$
\\
\hline
\end{tabular}
\end{center}
\vspace{-5mm} \caption{
$\mathbb{Z}_2$ projected characters of $D_N$ representations with odd $N$.}
\label{tab:Chara_odd_DN}
\end{table}

Let us study $n$-point interaction $\tilde \phi_{\mathbf{2}_{q_1}} \tilde \phi_{\mathbf{2}_{q_2}} \cdots \tilde \phi_{\mathbf{2}_{q_n}}$.
$q_i \in \{0, 1, \cdots ,N/2\}$ for even $N$ and $q_i \in \{0, 1, \cdots ,(N-1)/2\}$ for odd $N$.
When $q_i = 0$, $\tilde \phi_{\mathbf{2}_{q_i}}$ is interpreted as $\tilde{\mathbf{1}}_{++}$ or $\tilde{\mathbf{1}}_+$, 
and when $q_i = N/2$, $\tilde \phi_{\mathbf{2}_{q_i}}$ is interpreted as $\tilde{\mathbf{1}}_{+-}$.
The non-vanishing coupling condition Eq.~\eqref{eq:selection_rule_gauging} is calculated as
\begin{align}
	&\frac 1 {2N} \left[2 +  \sum_{m= 1}^{N/2  -1} |\tilde C_2^{(m)} | \cos \frac{2\pi q_1 m }{N} \cos \frac{2\pi q_2 m }{N} \cdots  \cos \frac{2\pi q_n m }{N} + 2(-1)^{q_1 + q_2 + \cdots +q_n}\right]
	\notag
	\\
	&= \frac{1}{2^{n-1}} \sum_{\epsilon_2, \cdots, \epsilon_n = \pm1 }\delta_{q_1 + \epsilon_2 q_2 + \cdots + \epsilon_n q_n, 0},
\end{align}
for even $N$ and 
\begin{align}
	&\frac 1 {2N} \left[2 +  \sum_{m= 1}^{\frac{N  -1} 2} | \tilde C_2^{(m)} | \cos \frac{2\pi q_1 m }{N} \cos \frac{2\pi q_2 m }{N} \cdots  \cos \frac{2\pi q_n m }{N} \right]
	\notag
	\\
	&= \frac{1}{2^{n-1}} \sum_{\epsilon_2, \cdots, \epsilon_n = \pm1 }\delta_{q_1 + \epsilon_2 q_2 + \cdots + \epsilon_n q_n, 0},
\end{align}
for odd $N$.
$n$-point interaction is allowed only when $q_1,q_2,  \cdots, q_n$ satisfies
\begin{align}
	q_1 + \epsilon_2 q_2 + \cdots + \epsilon_n q_n = 0 \quad \textrm{mod $N$}.
	\label{eq:selection_DN}
\end{align}
In terms of the $\mathbb{Z}_N$ symmetric model, 
there are $N/2 + 1$ $H$-invariant fields $\tilde{\phi}_q$ for even $N$ and $(N+1)/2$ fields for odd $N$.
There are two possibilities for the $\mathbb{Z}_2$ transformation of $\phi_{0, N/2}$. 
However, if they have a non-trivial transformation under $h$, namely $\phi_{0,N/2} \to -\phi_{0,N/2}$, these fields are projected out by the projection operator $P$ in Eq.~\eqref{eq:P}. 
These non-trivial transformations correspond to the representations $\mathbf{1}_{-+}$ and $\mathbf{1}_{--}$ for even $N$, and $\mathbf{1}_{-}$ for odd $N$. 
Thus, the remaining invariant field $\tilde \phi_{0}$ is interpreted as $\tilde{\mathbf{1}}_{++}$ or $\tilde{\mathbf{1}}_+$, and $\tilde \phi_{N/2}$ is interpreted as $\tilde{\mathbf{1}}_{+-}$. 
The other fields $\tilde \phi_q$ correspond to $\tilde{\mathbf{2}}_q$. 
Consequently, the non-invertible selection rule in Eq.~\eqref{eq:selection_rule} is reproduced by Eq.~\eqref{eq:selection_DN}.

\subsection*{Fusion-like algebra for $\mathbb{Z}_2$-gauged representations}

Here we calculate fusion-like algebra of $H$-invariant representations proposed in Section \ref{sec:Fusion-like Algebra}.
The CG coefficients among $H$-invariant fields are given by
\begin{align}
	(\tilde \phi_{\mathbf{1}_{++}}) (\tilde \phi_{\mathbf{1}_{++}}) &=  (\tilde \phi_{\mathbf{1}_{++}}),
	\quad
	(\tilde \phi_{\mathbf{1}_{++}}) (\tilde \phi_{\mathbf{1}_{+-}}) =  (\tilde \phi_{\mathbf{1}_{+-}}),
	\quad
	(\tilde \phi_{\mathbf{1}_{+-}}) (\tilde \phi_{\mathbf{1}_{+-}}) =  (\tilde \phi_{\mathbf{1}_{++}}),
	\notag
	\\
	(\tilde \phi_{\mathbf{1}_{++}}) (\tilde \phi_{\mathbf{2}_{q}}) &=  (\tilde \phi_{\mathbf{2}_{q}}),
	\quad
	(\tilde \phi_{\mathbf{1}_{+-}}) (\tilde \phi_{\mathbf{2}_{q}}) =  (\tilde \phi_{\mathbf{2}_{\frac N 2 - q}}),
	\notag
	\\
	(\tilde \phi_{\mathbf{2}_{q_1}}) (\tilde \phi_{\mathbf{2}_{q_2}}) &=  \frac 1 {\sqrt 2}(\tilde \phi_{\mathbf{2}_{q_1 + q_2}}) + \frac 1 {\sqrt 2}(\tilde \phi_{\mathbf{2}_{q_1 - q_2}}), \quad(q_1 - q_2 \neq 0, q_1 + q_2 \neq \frac N 2)
	\notag
	\\
	(\tilde \phi_{\mathbf{2}_{q_1}}) (\tilde \phi_{\mathbf{2}_{q_2}}) &=  \frac 1 {\sqrt 2} (\tilde \phi_{\mathbf{1}_{+-}}) + \frac 1 {\sqrt 2}  (\tilde \phi_{\mathbf{2}_{q_1  - q_2}}) , \quad(q_1 - q_2 \neq 0, q_1 + q_2 = \frac N 2)
	\notag
	\\
	(\tilde \phi_{\mathbf{2}_{q_1}}) (\tilde \phi_{\mathbf{2}_{q_2}}) &=  \frac 1 {\sqrt 2}(\tilde \phi_{\mathbf{2}_{q_1 + q_2}}) + \frac 1 {\sqrt 2}(\tilde \phi_{\mathbf{1}_{++}}), \quad(q_1 - q_2 = 0, q_1 + q_2 \neq \frac N 2)
	\notag
	\\
	(\tilde \phi_{\mathbf{2}_{q_1}}) (\tilde \phi_{\mathbf{2}_{q_2}}) &=  \frac 1 {\sqrt 2} (\tilde \phi_{\mathbf{1}_{++}}) + \frac 1 {\sqrt 2}  (\tilde \phi_{\mathbf{1}_{+-}}) , \quad(q_1 - q_2 = 0, q_1 + q_2 = \frac N 2)
\end{align}
for even $N$ and 
\begin{align}
	(\tilde\phi_{\mathbf{1}_{+}})  (\tilde\phi_{\mathbf{1}_{+}}) &=(\tilde\phi_{\mathbf{1}_{+}}),
	\quad
	(\tilde\phi_{\mathbf{1}_{+}})  (\tilde\phi_{\mathbf{2}_{q}}) = (\tilde\phi_{\mathbf{2}_{q}}),
	\notag
	\\
	(\tilde\phi_{\mathbf{2}_{q_1}})  (\tilde\phi_{\mathbf{2}_{q_2}}) &= \frac 1 {\sqrt 2} (\tilde\phi_{\mathbf{2}_{q_1 + q_2}}) + 
	 \frac 1 {\sqrt 2} (\tilde\phi_{\mathbf{2}_{q_1 - q_2}}), \quad (q_1 - q_2 \neq 0)
	\notag
	\\
	(\tilde\phi_{\mathbf{2}_{q_1}})  (\tilde\phi_{\mathbf{2}_{q_2}}) &= \frac 1 {\sqrt{2}}(\tilde\phi_{\mathbf{1}_{+}}) + \frac 1 {\sqrt 2} (\tilde\phi_{\mathbf{2}_{q_1 + q_2}})  \quad (q_1 - q_2 = 0)
\end{align}
for odd $N$.
On the other hand, $\tilde N^{\gamma}_{\alpha \beta}$ are given by 
\begin{align}
	\tilde{\mathbf{1}}_{++} \otimes \tilde{\mathbf{1}}_{++} &= \tilde{\mathbf{1}}_{++},
	\quad
	\tilde{\mathbf{1}}_{++} \otimes \tilde{\mathbf{1}}_{+-} = \tilde{\mathbf{1}}_{+-},
	\quad
	\tilde{\mathbf{1}}_{+-} \otimes \tilde{\mathbf{1}}_{+-} = \tilde{\mathbf{1}}_{++},
	\notag
	\\
	\tilde{\mathbf{1}}_{++} \otimes \tilde{\mathbf{2}}_{q} &= \frac 12 \tilde{\mathbf{2}}_{q},
	\quad
	\tilde{\mathbf{1}}_{+-} \otimes \tilde{\mathbf{2}}_{q} = \frac 12 \tilde{\mathbf{2}}_{\frac N 2 - q},
	\notag
	\\
	\tilde{\mathbf{2}}_{q_1} \otimes \tilde{\mathbf{2}}_{q_2} &= \frac 1 4 \tilde{\mathbf{2}}_{q_1 + q_2} + \frac 1 4 \tilde{\mathbf{2}}_{q_1 - q_2}, \quad (q_1 - q_2 \neq 0, q_1 + q_2 \neq \frac N 2)
	\notag
	\\
	\tilde{\mathbf{2}}_{q_1} \otimes \tilde{\mathbf{2}}_{q_2} &= \frac 1 2 \tilde{\mathbf{1}}_{+-} + \frac 1 4 \tilde{\mathbf{2}}_{q_1 - q_2}, \quad (q_1 - q_2 \neq 0, q_1 + q_2 = \frac N 2)
	\notag
	\\
	\tilde{\mathbf{2}}_{q_1} \otimes \tilde{\mathbf{2}}_{q_2} &= \frac 1 2 \tilde{\mathbf{1}}_{++} + \frac 1 4 \tilde{\mathbf{2}}_{q_1 + q_2}, \quad (q_1 - q_2 = 0, q_1 + q_2 \neq \frac N 2)
	\notag
	\\
	\tilde{\mathbf{2}}_{q_1} \otimes \tilde{\mathbf{2}}_{q_2} &= \frac 1 2 \tilde{\mathbf{1}}_{++} + \frac 1 2 \tilde{\mathbf{1}}_{+-}, \quad (q_1 - q_2 = 0, q_1 + q_2 = \frac N 2)
\end{align}
for even $N$.
For odd $N$ we obtain
\begin{align}
	\tilde{\mathbf{1}}_{+} \otimes \tilde{\mathbf{1}}_{+} &= \tilde{\mathbf{1}}_{+},
	\quad
	\tilde{\mathbf{1}}_{+} \otimes \tilde{\mathbf{2}}_{q} = \frac 1 2 \tilde{\mathbf{2}}_{q},
	\notag
	\\
	\tilde{\mathbf{2}}_{q_1} \otimes \tilde{\mathbf{2}}_{q_2} &= \frac 1 4 \tilde{\mathbf{2}}_{q_1 + q_2} + \frac 1 4 \tilde{\mathbf{2}}_{q_1 - q_2}, \quad (q_1 - q_2 \neq 0)
	\notag
	\\
	\tilde{\mathbf{2}}_{q_1} \otimes \tilde{\mathbf{2}}_{q_2} &= \frac 1 2 \tilde{\mathbf{1}}_{+} + \frac 1 4 \tilde{\mathbf{2}}_{q_1 + q_2}, \quad (q_1 - q_2 = 0).
	\label{eq:decomposition_odd_N}
\end{align}
These equations are consistent with Eq.~\eqref{eq:N_CG_relation}, and this fusion-like algebra correctly reproduce the irreducible decomposition of the product of $H$-invariant representations.
We also note that the tensor decomposition  of $\mathbb{Z}_2$-invariant representations are also consistent with the multiplication  rule for two orbit Eq.~\eqref{eq:fusion-like}.
Thus $\tilde N^{\gamma}_{\alpha\beta}$ has sufficient information to determine allowed bare couplings in this case.
For instance, Eq.~\eqref{eq:decomposition_odd_N} shows $\tilde{\mathbf{2}}_{q_1} \otimes \tilde{\mathbf{2}}_{q_2}  \otimes \tilde{\mathbf{2}}_{q_3}$ contains the trivial singlet only when $q_1 \pm q_2 \pm q_3 = 0$ is satisfied.

\subsection{$\mathbb{Z}_2$-gauging of $S_4\cong A_4 \rtimes \mathbb{Z}_2$}

Here we consider $\mathbb{Z}_2$-gauging of $S_4$ symmetry.
$S_4$ has 5 irreducible representations, and 10 independent field components.
After $\mathbb{Z}_2$-gauging, we obtain $\mathbb{Z}_2$-invariant representations given as
\begin{align}
\tilde{\Phi}_{\mathbf{1}} &= \tilde \phi_{\mathbf{1}} = \phi_{\mathbf{1}},
\hspace{5mm}
\tilde{\Phi}_{\mathbf{1}'} = (1-1) \Phi_{\mathbf{1}'} = 0,
\notag
\\
\tilde{\Phi}_{\mathbf{2}} &= \frac{1}{2}\left[
I_2 + 
\begin{pmatrix}
	0 & 1\\
	1 & 0
\end{pmatrix}
\right] \Phi_{\mathbf{2}} = 
\frac 1{2}
\begin{pmatrix}
(\phi_{\mathbf{2}})_1 + (\phi_{\mathbf{2}})_2\\
(\phi_{\mathbf{2}})_1 + (\phi_{\mathbf{2}})_2
\end{pmatrix}
= \tilde \phi_{\mathbf{2}} \ket{\tilde{\mathbf{2}}},
\notag
\\
\tilde{\Phi}_{\mathbf{3}} &= \frac{1}{2}\left[
I_3 + 
\begin{pmatrix}
	1 & 0 & 0\\
	0 & 0 & 1\\
	0 & 1 & 0
\end{pmatrix}
\right] \Phi_{\mathbf{3}} = 
\begin{pmatrix}
(\phi_{\mathbf{3}})_1\\
\frac{(\phi_{\mathbf{3}})_2 + (\phi_{\mathbf{3}})_3}2\\
\frac{(\phi_{\mathbf{3}})_2 + (\phi_{\mathbf{3}})_3}2\\
\end{pmatrix}= (\tilde\phi_{\mathbf{3}})_1 \ket{\tilde{\mathbf{3}}_1} + (\tilde\phi_{\mathbf{3}})_2 \ket{\tilde{\mathbf{3}}_2}
,
\notag
\\
\tilde{\Phi}_{\mathbf{3}'} &= \frac{1}{2}\left[
I_3 + 
\begin{pmatrix}
	-1 & 0 & 0\\
	0 & 0 & -1\\
	0 & -1 & 0
\end{pmatrix}
\right] \Phi_{\mathbf{3}'} = 
\begin{pmatrix}
0\\
\frac{(\phi_{\mathbf{3}'})_2 - (\phi_{\mathbf{3}'})_3}2\\
\frac{-(\phi_{\mathbf{3}'})_2 + (\phi_{\mathbf{3}'})_3}2\\
\end{pmatrix}
= \tilde \phi_{\mathbf{3}'} \ket{\tilde{\mathbf{3}}}.
\end{align}
$\ket{\tilde{\mathbf{r}}_i}$ denote a normalized basis where $P^{(\mathbf{r})}$ is diagonalized.
$\{ \ket{\tilde{\mathbf{3}}_{1}}, \ket{\tilde{\mathbf{3}}_{2} }\}$ is the orthonormal basis of $\mathbb{Z}_2$-invariant triplet.
We define
\begin{align}
	\ket{\tilde{\mathbf{3}}_{1}} = 
	\begin{pmatrix}
	1\\
	0\\
	0
	\end{pmatrix},
	&&
	\ket{\tilde{\mathbf{3}}_{2}} = 
	\frac 1 {\sqrt 2}\begin{pmatrix}
	0\\
	1\\
	1
	\end{pmatrix}.
\end{align}
We have 4 independent representations and 5 independent field components in this model.

$S_4$ is isomorphic to $A_4 \rtimes \mathbb{Z}_2 \cong ((\mathbb{Z}_2 \times \mathbb{Z}_2) \rtimes \mathbb{Z}_3) \rtimes \mathbb{Z}_2$.
Thus, its elements are written as 
\begin{align}
	a^i a'^j b^k h^l, 
\end{align}
where $a$ and $a'$ denote the $\mathbb{Z}_2$ generators that commute with each other, and $b$ denotes a generator of $\mathbb{Z}_3$. 
Here, $h$ denotes the $\mathbb{Z}_2$ generator that induces the outer automorphism of $A_4$. 
These generators satisfy $hah = a$, $ha'h = aa'$, and $hbh = b^2$. 
The double coset decomposition of $S_4$ with respect to $\mathbb{Z}_2$ is given by the sum of $\mathbb{Z}_2$-orbits of the right cosets $S_4/\mathbb{Z}_2$ under left multiplication by $\mathbb{Z}_2$. 
These double cosets are given by
\begin{align}
	\tilde{C}_1^{(1)} &= \{e, h\},
	\notag
	\\
	\tilde{C}_1^{(2)} &= \{a, ah\},
	\notag
	\\
	\tilde{C}_2^{(1)} &= \{a', aa', a'h, aa'h \},
	\notag
	\\
	\tilde{C}_2^{(2)} &= \{b, b^2, bh, b^2h \},
	\notag
	\\
	\tilde{C}_2^{(3)} &= \{a'b, aa'b^2, a'bh, aa'b^2h\},
	\notag
	\\
	\tilde{C}_2^{(4)} &= \{ab, ab^2, abh, ab^2h\},
	\notag
	\\
	\tilde{C}_2^{(5)} &= \{a a'b, a'b^2, a a'bh, a'b^2h \},
\end{align}
where the subscript indicates the number of elements contained in the double coset divided by $2$. 
The corresponding projected characters are summarized in Table~\ref{tab:Chara_S4}.
\begin{table}[htb!]
\begin{center}
\renewcommand{\arraystretch}{1.5} 
\setlength{\tabcolsep}{5pt} 
\begin{tabular}{| c |   c   c  c c |}
\hline
 & $\tilde{\chi}^{(\mathbf{1})}$ & $\tilde{\chi}^{(\mathbf{2})}$ & $\tilde{\chi}^{(\mathbf{3})}$ & $\tilde{\chi}^{(\mathbf{3}')}$\\ 
\hline
$\tilde{C}_1^{(1)}$ & $1$ &  $1$ &  $2$ &  $1$  
\\
$\tilde C_1^{(2)}$ & $1$ &  $1$ & 0 & $-1$
\\
$\tilde C_2^{(1)}$& $1$ &  $1$ & $-1$ & 0  
\\
$\tilde{C}_2^{(2)}$ & $1$ &  $-\frac 12$  & $\frac 12$ & $- \frac 12$  
\\
$\tilde{C}_2^{(3)}$ & $1$ &  $-\frac 12$  & $\frac 12$ & $-\frac 12$  
\\
$\tilde{C}_2^{(4)}$ & $1$ &  $-\frac 12$  & $-\frac 12$ & $\frac 12$  
\\
$\tilde{C}_2^{(5)}$ & $1$ &  $-\frac 12$  & $-\frac 12$ & $\frac 12$  
\\
\hline
\end{tabular}
\end{center}
\vspace{-5mm} \caption{
$\mathbb{Z}_2$ projected characters of $S_4$.
}
\label{tab:Chara_S4}
\end{table}

The condition in Eq.~\eqref{eq:selection_rule_gauging} for a non-vanishing $n$-point coupling 
$\tilde \phi_{\mathbf{1}}^k \tilde \phi_{\mathbf{2}}^l \tilde \phi_{\mathbf{3}}^m \tilde \phi_{\mathbf{3}'}^n$ is calculated from the table as
\begin{align}
	\frac{1}{12} \left[
	2^m + 0^m (-1)^n + 2 (-1)^m 0^n + 4\left(- \frac 12\right)^l \frac {(-1)^m + (-1)^n }{2^{m+n}}
	\right].
	\label{eq:general_form_S4}
\end{align}
Using the above expression, we explicitly determine the allowed bare couplings.

There are 15 independent 3-point couplings in $S_4$ invariant Lagrangians, which are given by 
\begin{align}
	& {\phi}_{\mathbf{1}}  {\phi}_{\mathbf{1}}  {\phi}_{\mathbf{1}},
	&&{\phi}_{\mathbf{1}} \phi_{\mathbf{1}'} \phi_{\mathbf{1}'},
	&&{\phi}_{\mathbf{1}}   {\phi}_{\mathbf{2}}  {\phi}_{\mathbf{2}},
	&&{\phi}_{\mathbf{1}}  {\phi}_{\mathbf{3}}  {\phi}_{\mathbf{3}},
	&&{\phi}_{\mathbf{1}}  {\phi}_{\mathbf{3}'}  {\phi}_{\mathbf{3}'},
	\notag
	\\
	& 
	{\phi}_{\mathbf{1}'}  \phi_{\mathbf{2}} \phi_{\mathbf{2}},
	&&{\phi}_{\mathbf{1}'} \phi_{\mathbf{3}}  \phi_{\mathbf{3}'},
	&&{\phi}_{\mathbf{2}} {\phi}_{\mathbf{2}}  {\phi}_{\mathbf{2}},
	&&{\phi}_{\mathbf{2}} {\phi}_{\mathbf{3}}  {\phi}_{\mathbf{3}},
	&&{\phi}_{\mathbf{2}}  {\phi}_{\mathbf{3}}  {\phi}_{\mathbf{3}'},
	\notag
	\\
	&
	{\phi}_{\mathbf{2}} {\phi}_{\mathbf{3}'}  {\phi}_{\mathbf{3}'},
	&&{\phi}_{\mathbf{3}}  {\phi}_{\mathbf{3}}  {\phi}_{\mathbf{3}},
	&&{\phi}_{\mathbf{3}}  {\phi}_{\mathbf{3}} {\phi}_{\mathbf{3}'},
	&&{\phi}_{\mathbf{3}}  {\phi}_{\mathbf{3}'}  {\phi}_{\mathbf{3}'},
	&&{\phi}_{\mathbf{3}'}  {\phi}_{\mathbf{3}'}  {\phi}_{\mathbf{3}'}.
\end{align}
After gauging, the fields are replaced by $\mathbb{Z}_2$-invariant fields.
As noted before, the non-trivial singlet $\phi_{\mathbf{1}'}$ is projected out by $\mathbb{Z}_2$-gauging.
Thus, any couplings containing $\phi_{\mathbf{1}'}$ also vanish.
Furthermore, the non-invertible selection rule in Eq.~\eqref{eq:general_form_S4} shows that the coupling $(\tilde{\phi}_{\mathbf{3}'})^3$ is eliminated by the $\mathbb{Z}_2$ projection, 
since 
\begin{align}
	\frac 1 {12} \left[
	1 + (-1)^3 + 4 \frac{1 + (-1)^3}{2^3}
	\right] = 0.
\end{align}
This reflects the fact that the three-point coupling $(\phi_{\mathbf{3}})^3$ does not appear in the $A_4$ model when $U=U_-$ \footnote{See Section~\ref{sec:A_4}.}. 
As a result, $11$ independent bare couplings remain from the original Lagrangian.

For the 4-point couplings, there are $40$ independent couplings in the original Lagrangian.
After gauging, $23$ independent bare couplings remain.
Among the 40 initial interactions, 15 contains $\phi_{\mathbf{1}'}$ and are trivially eliminated by the $\mathbb{Z}_2$ projection.
The interactions that are non-trivially prohibited by the $\mathbb{Z}_2$-gauging are the following two couplings
\begin{align}
	\tilde{\phi}_{\mathbf{1}} \tilde{\phi}_{\mathbf{3}'} \tilde{\phi}_{\mathbf{3}'} \tilde{\phi}_{\mathbf{3}'},
	&&
	\tilde{\phi}_{\mathbf{2}} \tilde{\phi}_{\mathbf{3}'} \tilde{\phi}_{\mathbf{3}'} \tilde{\phi}_{\mathbf{3}'}.
\end{align}
The former is prohibited since $\tilde{\mathbf{3}}' \otimes \tilde{\mathbf{3}}'$ does not contain $\tilde{\mathbf{3}}'$, thus they cannot form the trivial singlet.
The latter implies that $\tilde{\mathbf{3}}' \otimes \tilde{\mathbf{3}}' \otimes \tilde{\mathbf{3}}'$ does not contain $\tilde{\mathbf{2}}$, although $\tilde{\mathbf{3}}' \otimes \tilde{\mathbf{3}}'$ contains $\tilde{\mathbf{3}}$, and $\tilde{\mathbf{3}} \otimes \tilde{\mathbf{3}}'$ contains $\tilde{\mathbf{2}}$.
This fact cannot be read off from the $S_4$ representation algebra alone.
However, it is obtained by the non-invertible selection rule based on characters, since
\begin{align}
	\frac 1 {12} \left[
	1 + (-1)^3 + 4 \left(-\frac{1}2\right) \frac{1 + (-1)^3}{2^3}
	\right] = 0,
\end{align}
which shows that $\tilde{\phi}_{{\mathbf{2}}}(\tilde{\phi}_{{\mathbf{3}}'})^3$ does not contain the trivial singlet.

We summarize the non-trivial interactions which are prohibited by the $\mathbb{Z}_2$-gauging up to 7-point couplings in Table \ref{tab:int_S4_orbi}.
The table shows that bare couplings composed exclusively of $\tilde{\mathbf{2}}$ fields and an odd number of $\tilde{\mathbf{3}}'$ fields are prohibited by the $\mathbb{Z}_2$ projection.
This rule can be derived from the non-invertible selection rule base on characters.
When $m = 0$  and $n$ is odd,  Eq.~\eqref{eq:general_form_S4} is calculated as  
\begin{align}
	\frac 1 {12} \left[
	1 + (-1)^{n} + 4 \left(-\frac{1}2\right)^l \frac{1 + (-1)^n}{2^n}
	\right] = 0,
\end{align}
and hence, such bare couplings are projected out by the $\mathbb{Z}_2$-projection.
\begin{table}[htb!]
\begin{center}
\renewcommand{\arraystretch}{1.5} 
\setlength{\tabcolsep}{5pt} 
\begin{tabular}{| c | c | }
\hline
$n$-point & prohibited bare interactions \\ 
\hline
3 & $\tilde{\phi}_{\mathbf{3}'}^3$
\\
4 & $\tilde{\phi}_{\mathbf{1}}\tilde{\phi}_{\mathbf{3}'}^3, \tilde{\phi}_{\mathbf{2}}\tilde{\phi}_{\mathbf{3}'}^3$
\\
5 & $\tilde{\phi}_{\mathbf{1}}^2\tilde{\phi}_{\mathbf{3}'}^3, \tilde{\phi}_{\mathbf{1}} \tilde{\phi}_{\mathbf{2}}\tilde{\phi}_{\mathbf{3}'}^3,  \tilde{\phi}_{\mathbf{2}}^2\tilde{\phi}_{\mathbf{3}'}^3,  \tilde{\phi}_{\mathbf{3}'}^5$
\\
6 & 
$\tilde{\phi}_{\mathbf{1}}^3\tilde{\phi}_{\mathbf{3}'}^3,  \tilde{\phi}_{\mathbf{1}} \tilde{\phi}_{\mathbf{2}}^2\tilde{\phi}_{\mathbf{3}'}^3,  \tilde{\phi}_{\mathbf{1}} \tilde{\phi}_{\mathbf{3}'}^5, \tilde{\phi}_{\mathbf{1}}^2 \tilde{\phi}_{\mathbf{2}}\tilde{\phi}_{\mathbf{3}'}^3, \tilde{\phi}_{\mathbf{2}}^3\tilde{\phi}_{\mathbf{3}'}^3,  \tilde{\phi}_{\mathbf{2}} \tilde{\phi}_{\mathbf{3}'}^5$
\\
7 & 
$
\tilde{\phi}_{\mathbf{1}}^4 \tilde{\phi}_{\mathbf{3}'}^3, 
\tilde{\phi}_{\mathbf{1}}^2 \tilde{\phi}_{\mathbf{2}}^2\tilde{\phi}_{\mathbf{3}'}^3, 
\tilde{\phi}_{\mathbf{1}}^2 \tilde{\phi}_{\mathbf{3}'}^5,
\tilde{\phi}_{\mathbf{1}}^3 \tilde{\phi}_{\mathbf{2}}\tilde{\phi}_{\mathbf{3}'}^3,
\tilde{\phi}_{\mathbf{1}} \tilde{\phi}_{\mathbf{2}}^3\tilde{\phi}_{\mathbf{3}'}^3, 
\tilde{\phi}_{\mathbf{1}} \tilde{\phi}_{\mathbf{2}} \tilde{\phi}_{\mathbf{3}'}^5,  
\tilde{\phi}_{\mathbf{2}}^2 \tilde{\phi}_{\mathbf{3}'}^5,
\tilde{\phi}_{\mathbf{2}}^4 \tilde{\phi}_{\mathbf{3}'}^3,
\tilde{\phi}_{\mathbf{3}'}^7
$
\\
\hline
\end{tabular}
\end{center}
\vspace{-5mm} \caption{Prohibited bare interactions in the $\mathbb{Z}_2$ gauged $S_4$ symmetric models.
We omit interactions including $\tilde{\phi}_{\mathbf{1}'}$ since it is trivial.}
\label{tab:int_S4_orbi}
\end{table}

\subsection*{Fusion-like algebra for $\mathbb{Z}_2$-invariant representations}

Here we calculate fusion-like algebra of $H$-invariant representations.
The CG coefficients among $H$-invariant fields are summarized in Table \ref{tab:S4}.
\begin{table}[htb!]
\begin{center}
\renewcommand{\arraystretch}{1.35} 
\setlength{\tabcolsep}{5pt} 
{\small 
\begin{tabular}{| c | c | c | c | c  | c | }
\hline
 & $\tilde \phi_{\mathbf{1}}$  & $\tilde \phi_{\mathbf{2}}$  & $(\tilde \phi_{\mathbf{3}})_{1}$ & $(\tilde \phi_{\mathbf{3}})_{2} $ & $\tilde \phi_{\mathbf{3}'} $ \\ 
\hline
$\tilde \phi_{\mathbf{1}}$ &  $\tilde \phi_{\mathbf{1}}$ &  $\tilde \phi_{\mathbf{2}}$& $(\tilde \phi_{\mathbf{3}})_{1}$ & $(\tilde \phi_{\mathbf{3}})_{2} $  & $\tilde \phi_{\mathbf{3}'} $\\ 
$\tilde \phi_{\mathbf{2}}$ & $\tilde \phi_{\mathbf{2}}$  & $\frac 1{\sqrt 2}  \tilde \phi_{\mathbf{1}} +  \frac 1{\sqrt 2} \tilde \phi_{\mathbf{2}}$ & $(\tilde \phi_{\mathbf{3}})_{1}$ & $- \frac 12 (\tilde \phi_{\mathbf{3}})_{2} - \frac {\sqrt 3 i}2 \tilde \phi_{\mathbf{3}'}$  & $- \frac 1 2 \tilde \phi_{\mathbf{3}'}-\frac{\sqrt 3}2  i  (\tilde \phi_\mathbf{3})_2$\\
$(\tilde \phi_{\mathbf{3}})_{1}$ & $(\tilde \phi_{\mathbf{3}})_1$  & $(\tilde \phi_{\mathbf{3}})_{1}$  & $\frac 1{\sqrt 3} \tilde \phi_{\mathbf{1}} + \sqrt {\frac 2 3} \tilde \phi_{\mathbf{2}}$ & $\frac 1 {\sqrt 2} (\tilde \phi_{\mathbf{3}})_{2} -  \frac 1 {\sqrt 2} \tilde \phi_{\mathbf{3}'}$ & $-\frac 1{\sqrt 2} \tilde{\phi}_{\mathbf{3}'} + \frac 1{\sqrt 2} (\tilde{\phi}_{\mathbf{3}})_2$ \\
$(\tilde \phi_{\mathbf{3}})_{2} $ & $(\tilde \phi_{\mathbf{3}})_2$  & $- \frac 12 (\tilde \phi_{\mathbf{3}})_{2}  - \frac{\sqrt 3 i}2  \tilde \phi_{\mathbf{3}'}$   & $\frac 1 {\sqrt 2} (\tilde \phi_{\mathbf{3}})_{2} +  \frac 1 {\sqrt 2} \tilde \phi_{\mathbf{3}'}$ &   $ -\frac 1{\sqrt 3} \tilde \phi_{\mathbf{1}}  - \frac 1 {\sqrt{6}} \tilde \phi_{\mathbf{2}} + \frac 1{\sqrt 2} (\tilde \phi_{\mathbf{3}})_{1}$ & $\frac {i}{\sqrt 2} \tilde \phi_{\mathbf{2}} - \frac 1 {\sqrt 2} (\tilde \phi_{\mathbf{3}})_1$ \\
$(\tilde \phi_{\mathbf{3}'}) $ & $\tilde \phi_{\mathbf{3}'}$  & $- \frac 1 2 \tilde \phi_{\mathbf{3}'}- \frac{\sqrt 3}2 i  (\tilde \phi_\mathbf{3})_2$   & $-\frac 1{\sqrt 2} \tilde{\phi}_{\mathbf{3}'} + \frac 1{\sqrt 2} (\tilde{\phi}_{\mathbf{3}})_2$ &   $\frac {i}{\sqrt 2} \tilde \phi_{\mathbf{2}} - \frac 1 {\sqrt 2} (\tilde \phi_{\mathbf{3}})_1$ & $\frac 1{\sqrt 3} \tilde \phi_{\mathbf{1}} - \frac 1 {\sqrt 6} \tilde \phi_{\mathbf 2} -\frac 1 {\sqrt 2} (\tilde \phi_{\mathbf{3}})_1$\\
\hline
\end{tabular}
}
\end{center}
\vspace{-5mm} \caption{Multiplication rules for $\mathbb{Z}_2$-projected $S_4$ fields. }
\label{tab:S4}
\end{table}

$\tilde N^\gamma_{\alpha \beta}$ is also calculated as
\begin{align}
	\tilde{\mathbf{1}} \otimes \tilde{\mathbf{1}} &= \tilde{\mathbf{1}},
	&&
	\tilde{\mathbf{1}} \otimes \tilde{\mathbf{2}} = \frac 12 \tilde{\mathbf{2}},
	&&
	\tilde{\mathbf{1}} \otimes \tilde{\mathbf{3}} = \frac 23 \tilde{\mathbf{3}},
	\notag
	\\
	\tilde{\mathbf{1}} \otimes \tilde{\mathbf{3}}' &= \frac 13 \tilde{\mathbf{3}}',
	\notag
	\\
	\tilde{\mathbf{2}} \otimes  \tilde{\mathbf{2}} &= \frac 12 \tilde{\mathbf{1}} +  \frac 14 \tilde{\mathbf{2}},
	&&
	\tilde{\mathbf{2}} \otimes  \tilde{\mathbf{3}} =  \frac {5}{12} \tilde{\mathbf{3}} + \frac{1}{4} \tilde{\mathbf{3}}',
	&&
	\tilde{\mathbf{2}} \otimes  \tilde{\mathbf{3}}' = \frac14 \tilde{\mathbf{3}} + \frac 1{12} \tilde{\mathbf{3}}',
	\notag
	\\
	\tilde{\mathbf{3}} \otimes  \tilde{\mathbf{3}} &= \frac{2}{3} \tilde{\mathbf{1}} + \frac{5}{12} \tilde{\mathbf{2}} + \frac{1}{2} \tilde{\mathbf{3}} + \frac{1}{3} \tilde{\mathbf{3}}',
	&&
	\tilde{\mathbf{3}} \otimes  \tilde{\mathbf{3}}' = \frac{1}{4} \tilde{\mathbf{2}} + \frac13\tilde{\mathbf{3}} + \frac16\tilde{\mathbf{3}}',
	&&
	\tilde{\mathbf{3}}' \otimes  \tilde{\mathbf{3}}' = \frac{1}{3} \tilde{\mathbf{1}} + \frac1{12}\tilde{\mathbf{2}} + \frac{1}{6} \tilde{\mathbf{3}}
	.
	\label{eq:N^g_ab_S}
\end{align}
$\tilde{\mathbf{3}}'$ does not appear in  $\tilde{\mathbf{3}}' \otimes  \tilde{\mathbf{3}}'$, and hence a cubic term $(\tilde{\phi}_{\mathbf{3}'})^3 $ is projected out in the $\mathbb{Z}_2$-gauged theory at the tree level.
This fact is consistent with the allowed bare couplings calculated by the projected character.

\subsection{$\mathbb{Z}_2$ gauging of $\Delta(54) \cong \Delta(27) \rtimes \mathbb{Z}_2$}

$\Delta(54)$ is a discrete group of order $54$, which is isomorphic to $((\mathbb{Z}_3 \times \mathbb{Z}_3) \rtimes \mathbb{Z}_3) \rtimes \mathbb{Z}_2 \cong \Delta(27) \rtimes \mathbb{Z}_2$.
Hence, all the elements of $\Delta(54)$ can be expressed as $g = a^m a'^n b^k h^l$,
where $a,a',b$ are order-3 generators and $h$ is the $\mathbb{Z}_2$ generator, i.e., $a^3 = a'^3 = b^3 = h^2 = e$.
They satisfy $hah = a'^2, ha'h = a^2, bab^{-1} = a^2 a'^2, ba'b^{-1} = a$, and $hbh = b^2$.
There are 10 conjugacy classes, and hence, we have 10 irreducible representations:
2 singlets $\mathbf{1}$ and $\mathbf{1}'$, 4 real doublets $\mathbf{2}_k$, and 4 complex triplets $\mathbf{3}_{ij}$.
Here, $i,j$ take the value of 1 or 2.
$\mathbf{3}_{i1}^* = \mathbf{3}_{i2}$.
We consider $\mathbb{Z}_2$-gauging, which is generated by $h$.
After gauging, $\mathbb{Z}_2$-invariant fields are given by 
\begin{align}
	\tilde{\Phi}_{\bm{1}} &= \tilde{\phi}_{\bm{1}} 
	=  {\phi}_{\bm{1}},
	\notag
	\\
	\tilde{\Phi}_{\bm{1}'} &= \frac12(1-1) {\Phi}_{\bm{1}'} = 0
	\notag
	\\
	\tilde{\Phi}_{\bm{2}_k} &= \frac12 \left[
	I_{2} + \begin{pmatrix}
	0 & 1\\
	1 &0
	\end{pmatrix}\right] {\Phi}_{\bm{2}_k}
	= \frac 12
	\begin{pmatrix}
	(\phi_{\mathbf{2}_k})_1+(\phi_{\mathbf{2}_k})_2\\
	(\phi_{\mathbf{2}_k})_1+(\phi_{\mathbf{2}_k})_2
	\end{pmatrix}
	= \tilde{\phi}_{\bm{2}_k} \ket{\tilde{\mathbf{2}}_k},
	\notag
	\\
	\tilde{\Phi}_{\bm{3}_{1k}} &= \frac 12 \left[ I_{3}+ 
	\begin{pmatrix}
	0 & 0 & 1\\
	0 & 1 & 0\\
	1 &0 &0
	\end{pmatrix}
	\right]{\Phi}_{\bm{3}_{1k}}
	= \begin{pmatrix}
	\frac{({\phi}_{\mathbf{3}_{1k}})_1 +({\phi}_{\mathbf{3}_{1k}})_3}{2}\\
	({\phi}_{\mathbf{3}_{1k}})_2 \\
	\frac{({\phi}_{\mathbf{3}_{1k}})_1 +({\phi}_{\mathbf{3}_{1k}})_3}{2}\\
	\end{pmatrix}
	= (\tilde {\phi}_{\mathbf{3}_{1k}})_1 \ket{(\tilde{\mathbf{3}}_{1k})_1} + (\tilde {\phi}_{\mathbf{3}_{1k}})_2 \ket{(\tilde{\mathbf{3}}_{1k})_2},
	\notag
	\\
	\tilde{\Phi}_{\bm{3}_{2k}} &= \frac12 \left[ I_3 + 
	\begin{pmatrix}
	0 & 0 & -1\\
	0 & -1 & 0\\
	-1 &0 &0
	\end{pmatrix}\right]{\Phi}_{\bm{3}_{2k}}
	= 
	\begin{pmatrix}
	\frac{({\phi}_{\mathbf{3}_{1k}})_1 - ({\phi}_{\mathbf{3}_{1k}})_3}{2}\\
	0\\
	-\frac{({\phi}_{\mathbf{3}_{1k}})_1 - ({\phi}_{\mathbf{3}_{1k}})_3}{2} 
	\end{pmatrix} = \tilde{\phi}_{\bm{3}_{2k}} \ket{\tilde{\mathbf{3}}_{2k}},
\end{align}
where $\ket{{\mathbf{2}}_k}, \ket{(\tilde{\mathbf{3}}_{1k})_{1}}, \ket{(\tilde{\mathbf{3}}_{1k})_{2}} \}$ and $\ket{(\tilde{\mathbf{3}}_{2k})}$ are the orthonormal basis vectors of $\mathbb{Z}_2$-invariant subspaces.
We define
\begin{align}
	\ket{\mathbf{2}_k} = 
	\frac 1 {\sqrt 2}\begin{pmatrix}
	1\\
	1
	\end{pmatrix},
	&&
	\ket{(\tilde{\mathbf{3}}_{1k})_{1}} = 
	\frac 1 {\sqrt 2}\begin{pmatrix}
	1\\
	0\\
	1
	\end{pmatrix},
	&&
	\ket{(\tilde{\mathbf{3}}_{1k})_{2}} = 
	\begin{pmatrix}
	0\\
	1\\
	0
	\end{pmatrix},
	&&
	\ket{(\tilde{\mathbf{3}}_{2k})_{2}} = 
	\frac 1 {\sqrt 2}\begin{pmatrix}
	1\\
	0\\
	-1
	\end{pmatrix}.
\end{align}
In the original theory, we have 22 independent degrees of freedom, but there remain only 11 contents after the $\mathbb{Z}_2$-projection.

The double coset decomposition of $\Delta(54)$ in terms of $\mathbb{Z}_2$ is given by
\begin{align}
	\tilde C_1^{(1)} &= \{e, h \},
	&&
	\tilde C_1^{(2)} = \{a a'^2, a a'^2 h \},
	\notag
	\\
	\tilde C_1^{(3)} &= \{a^2 a', a^2 a' h \},
	&&
	\tilde C_2^{(1)} = \{a, a'^2, ah, a'^2h \}, 
	\notag
	\\
	\tilde C_2^{(2)} &= \{a', a^2, a'h, a^2h \} ,
	&&
	\tilde C_2^{(3)} = \{aa', a^2a'^2, aa'h, a^2a'^2h \} ,
	\notag
	\\
	\tilde C_2^{(4)} &= \{b, b^2, bh, b^2h \},
	&&
	\tilde C_2^{(5)} = \{ab, a'^2b^2, abh, a'^2b^2 h\},
	\notag
	\\
	\tilde C_2^{(6)} &= \{a^2b, a'b^2, a^2bh, a'b^2 h\} ,
	&&
	\tilde C_2^{(7)} = \{a' b, a^2b^2, a' bh, a^2b^2 h\} ,
	\notag
	\\
	\tilde C_2^{(8)} &= \{a'^2b, ab^2, a'^2 bh, ab^2 h\} ,
	&&
	\tilde C_2^{(9)} = \{aa' b, a^2a'^2b^2, aa' bh, a^2a'^2b^2h\},
	\notag
	\\
	\tilde C_2^{(10)} &= \{aa'^2 b, aa'^2b^2, aa'^2 bh, aa'^2b^2h\},
	&&
	\tilde C_2^{(11)} = \{a^2a' b, a^2 a'b^2, a^2a' bH, a^2 a'b^2h\}, 
	\notag
	\\
	C_2^{(12)} &= \{a^2a'^2 b, a a'b^2, a^2a'^2 bh, a a'b^2h\}.
\end{align}
The projected characters are shown in Table \ref{tab:Chara_Delta54}.
The non-vanishing coupling condition Eq.~\eqref{eq:selection_rule_gauging} for $n$-point couplings is calculated from the table.
Using it, we explicitly determine the allowed bare couplings.
\begin{table}[htb!]
\begin{center}
\renewcommand{\arraystretch}{1.5} 
\setlength{\tabcolsep}{5pt} 
\begin{tabular}{| c |   c   c c c c c  c|}
\hline
 & $\tilde{\chi}^{(\mathbf{1})}$ 
 & $\tilde{\chi}^{(\mathbf{2}_1)}$ & $\tilde{\chi}^{(\mathbf{2}_2)}$ & $\tilde{\chi}^{(\mathbf{2}_3)}$  & $\tilde{\chi}^{(\mathbf{2}_4)}$  
 & $\tilde{\chi}^{(\mathbf{3})_{1k}}$ 
 & $\tilde{\chi}^{(\mathbf{3})_{2k}}$
 \\ 
\hline
$\tilde{C}_1^{(1)}$ & $1$ &  $1$ &  $1$ &  $1$ & $1$ & $2$  & $1$
\\
$\tilde C_1^{(2)}$ & $1$ &  $1$  & $1$ & $1$ & 1
& $2\omega^k $ & $\omega^k $ 
\\
$\tilde C_1^{(3)}$ & $1$ &  $1$ & $1$ & $1$ & $1$ 
& $2\omega^{2k}$ 
& $\omega^{2k}$
\\
$\tilde C_2^{(1)}$& $1$ &  $1$ & $- \frac 12$ & $ -\frac 12$ & $-\frac 12$ 
& $\frac {\omega^{2k}}2 $ 
& $-\frac {\omega^{2k}}2  $
\\
$\tilde C_2^{(2)}$& $1$ &  $1$ & $- \frac 12$ & $ -\frac 12$  & $-\frac 12$ 
& $\frac {\omega^k} 2$ 
& $- \frac {\omega^k} 2 $
\\
$\tilde C_2^{(3)}$& $1$ &  $1$ & $-\frac 12$ & $ -\frac 12$ &  $-\frac 12 $ 
& $\frac 12$ 
& $-\frac 12$
\\
$\tilde C_2^{(4)}$& $1$ &  $-\frac 12$ & $-\frac 12 $ & $ -\frac 12$ & $1$  
& $\frac 12$ 
& $- \frac 12$
\\
$\tilde C_2^{(5)}$& $1$ &  $-\frac 12$ & $1$ & $ -\frac 12$ & $-\frac 12$ 
& $\frac 12$
& $-\frac 12$
\\
$\tilde C_2^{(6)}$& $1$ &  $-\frac 12$ & $-\frac 12$ & $1$ &  $-\frac 12$ 
& $\frac 12$ 
& $-\frac 12$
\\
$\tilde C_2^{(7)}$& $1$ &  $-\frac 12$ & $1$ & $ -\frac 12$ & $-\frac 12$ 
& $\frac{\omega^{2k}}2  $ 
& $-\frac{\omega^{2k}}2  $
\\
$\tilde C_2^{(8)}$& $1$ &  $-\frac 12$ & $-\frac 12$ & $1$ & $-\frac 12$ 
& $\frac{\omega^k}2  $ 
& $-\frac{\omega^k}2 $
\\
$\tilde C_2^{(9)}$& $1$ &  $-\frac 12$ & $-\frac 12$ & $1$ & $-\frac 12$ 
& $\frac{\omega^{2k}}2  $
& $-\frac{\omega^{2k}}2   $ 
\\
$\tilde C_2^{(10)}$& $1$ &  $-\frac 12$ & $-\frac 12$ & $ -\frac 12$ &  $1$ 
& $\frac{\omega^k}2  $ 
& $-\frac{\omega^k}2 $ 
\\
$\tilde C_2^{11)}$& $1$ &  $-\frac 12$ & $-\frac 12$ & $ -\frac 12$ & $1$ 
& $\frac{\omega^{2k}}2 $ 
& $-\frac{\omega^{2k}}2  $ 
\\
$\tilde C_2^{(12)}$& $1$ &  $-\frac 12$ & $1$ & $ -\frac 12$ & $-\frac 12$ 
& $\frac{\omega^k}2  $ 
& $-\frac{\omega^k}2 $ 
\\
\hline
\end{tabular}
\end{center}
\vspace{-5mm} \caption{
$\mathbb{Z}_2$ projected characters of $\Delta(54)$.
}
\label{tab:Chara_Delta54}
\end{table}

There are 46 independent 3-point couplings before gauging.
After gauging, there remain 37 bare couplings, and the following $9$ interaction terms are eliminated by the $\mathbb{Z}_2$ projection:
\begin{align}
	& \tilde{\phi}_{\mathbf{1}}   \tilde{\phi}_{\mathbf{1}'}   \tilde{\phi}_{\mathbf{1}'},
	&&\tilde{\phi}_{\mathbf{1}'}  \tilde\phi_{{\mathbf{2}}_1}  \tilde\phi_{{\mathbf{2}}_1},
	&&\tilde{\phi}_{\mathbf{1}'}  \tilde\phi_{{\mathbf{2}}_2}  \tilde\phi_{{\mathbf{2}}_2},
	\notag
	\\
	&\tilde{\phi}_{\mathbf{1}'}  \tilde\phi_{{\mathbf{2}}_3}  \tilde\phi_{{\mathbf{2}}_3},
	&&\tilde{\phi}_{\mathbf{1}'}  \tilde\phi_{{\mathbf{2}}_4}  \tilde\phi_{{\mathbf{2}}_4},
	&&\tilde{\phi}_{\mathbf{1}'}  \tilde\phi_{\mathbf{3}_{11}}  \tilde\phi_{\mathbf{3}_{22}},
	\notag
	\\
	&\tilde{\phi}_{\mathbf{1}'}  \tilde\phi_{\mathbf{3}_{12}}  \tilde\phi_{\mathbf{3}_{21}},
	&&\tilde{\phi}_{\mathbf{3}_{21}}  \tilde\phi_{\mathbf{3}_{21}}  \tilde\phi_{\mathbf{3}_{21}},
	&&\tilde{\phi}_{\mathbf{3}_{22}}  \tilde\phi_{\mathbf{3}_{22}}  \tilde\phi_{\mathbf{3}_{22}}.
\end{align}
The $\mathbb{Z}_2$-gauging deletes $\phi_{\mathbf{1}'}$ since it is $\mathbb{Z}_2$-odd.
Thus, $(\tilde{\phi}_{\mathbf{3}_{2k}})^3$ are the only bare interactions which are non-trivially prohibited by the $\mathbb{Z}_2$-gauging of the $\Delta(54)$ symmetry.
In fact, the non-vanishing condition for $(\tilde \phi_{\mathbf{3}_{2k}})^3$ is calculated as
\begin{align}
	\frac{2}{54}\left[
	1+ \omega^{3k} + \omega^{6k} + 2\left( 4 \frac{(-1)^3 + (-\omega^{k})^3 + (-\omega^{2k})^3}{2^3}
	\right)
	\right] = 0,
	\label{eq:projected_phi^3}
\end{align}
and hence the trivial singlet in $\tilde{\mathbf{3}}_{2k} \otimes \tilde{\mathbf{3}}_{2k} \otimes \tilde{\mathbf{3}}_{2k}$ is projected out.

We summarize the non-trivial bare interactions which are prohibited by the $\mathbb{Z}_2$-gauging up to 7-point couplings in Table \ref{tab:int_Delta54_orbi}.
It shows that the allowed bare couplings of the $\Delta(54)$ symmetry with $\mathbb{Z}_2$-gauging are identical to those of the model without the $\mathbb{Z}_2$-projection, with the exception of a few specific terms.

\begin{table}[htb!]
\begin{center}
\renewcommand{\arraystretch}{1.5} 
\setlength{\tabcolsep}{5pt} 
\begin{tabular}{| c | c | }
\hline
$n$-point & prohibited bare interactions \\ 
\hline
3 & $\tilde{\phi}_{\mathbf{3}_{21}}^3, \tilde{\phi}_{\mathbf{3}_{22}}^3$
\\
4 & 
$
\tilde{\phi}_{\mathbf{1}} \tilde{\phi}_{\mathbf{3}_{21}}^3, 
\tilde{\phi}_{\mathbf{1}} \tilde{\phi}_{\mathbf{3}_{22}}^3
$
\\
5 & 
$
\tilde{\phi}_{\mathbf{1}}^2 \tilde{\phi}_{\mathbf{3}_{21}}^3, 
\tilde{\phi}_{\mathbf{1}}^2 \tilde{\phi}_{\mathbf{3}_{22}}^3
$
\\
6 & 
$
\tilde{\phi}_{\mathbf{1}}^3 \tilde{\phi}_{\mathbf{3}_{21}}^3, 
\tilde{\phi}_{\mathbf{1}}^3 \tilde{\phi}_{\mathbf{3}_{22}}^3
$
\\
7 &
$
\tilde{\phi}_{\mathbf{1}}^4 \tilde{\phi}_{\mathbf{3}_{21}}^3, 
\tilde{\phi}_{\mathbf{1}}^4 \tilde{\phi}_{\mathbf{3}_{22}}^3
$
\\
\hline
\end{tabular}
\end{center}
\vspace{-5mm} \caption{
Prohibited bare interactions in the $\mathbb{Z}_2$-gauged $\Delta(54)$ symmetric models.
We omit interactions containing $\tilde{\phi}_{\mathbf{1}'}$ since they trivially vanish.}
\label{tab:int_Delta54_orbi}
\end{table}

\subsubsection*{Fusion-like algebra for $\mathbb{Z}_2$-invariant representations}

The CG coefficients are summarized in Appendix \ref{sec:CG_D54}.
After $\mathbb{Z}_2$-gauging, we obtain
\begin{align}
	&\tilde{\mathbf{1}} \otimes {}^{\forall}\tilde{\mathbf{r}} = \frac{\dim \tilde V^{(\mathbf{r})}}{\dim V^{(\mathbf{r})}} \tilde{\mathbf{r}},
	&&
	\tilde{\mathbf{2}}_i \otimes \tilde{\mathbf{2}}_i = \frac 12 \tilde{\mathbf{1}} + \frac 14 \tilde{\mathbf{2}}_i, 
	&&
	\tilde{\mathbf{2}}_i \otimes \tilde{\mathbf{2}}_{j} = \frac 14 \tilde{\mathbf{2}}_{k} + \frac 14 \tilde{\mathbf{2}}_{l},
	\notag
	\\
	&
	\tilde{\mathbf{2}}_i\otimes \tilde{\mathbf{3}}_{1j} = \frac{5}{12}\tilde{\mathbf{3}}_{1j} +\frac{1}{4}\tilde{\mathbf{3}}_{2j},
	&&
	\tilde{\mathbf{2}}_i\otimes \tilde{\mathbf{3}}_{2j} = \frac{1}{4}\tilde{\mathbf{3}}_{1j} +\frac{1}{12}\tilde{\mathbf{3}}_{2j},
	&&
	\tilde{\mathbf{3}}_{11} \otimes \tilde{\mathbf{3}}_{11} = \tilde{\mathbf{3}}_{12} +\frac{1}{3}\tilde{\mathbf{3}}_{22},
	\notag
	\\
	&
	\tilde{\mathbf{3}}_{12} \otimes \tilde{\mathbf{3}}_{12} = \tilde{\mathbf{3}}_{11} +\frac{1}{3}\tilde{\mathbf{3}}_{21},
	&&
	\tilde{\mathbf{3}}_{21} \otimes \tilde{\mathbf{3}}_{21} = \frac{1}{3}\tilde{\mathbf{3}}_{12},
	&&
	\tilde{\mathbf{3}}_{22} \otimes \tilde{\mathbf{3}}_{22} = \frac{1}{3}\tilde{\mathbf{3}}_{11},
	\notag
	\\
	&
	\tilde{\mathbf{3}}_{11} \otimes \tilde{\mathbf{3}}_{12} = \frac{2}{3}\tilde{\mathbf{1}} + \frac{5}{12} \sum_{i= 1}^4 \tilde{\mathbf{2}}_i,
	&&
	\tilde{\mathbf{3}}_{21} \otimes \tilde{\mathbf{3}}_{22} = \frac{1}{3}\tilde{\mathbf{1}} + \frac{1}{12} \sum_{i= 1}^4 \tilde{\mathbf{2}}_i,
	&&
	\tilde{\mathbf{3}}_{12} \otimes \tilde{\mathbf{3}}_{21} =\frac{1}{4} \sum_{i= 1}^4 \tilde{\mathbf{2}}_i,
	\notag
	\\
	&
	\tilde{\mathbf{3}}_{11} \otimes \tilde{\mathbf{3}}_{22} =  \frac{1}{4} \sum_{i= 1}^4 \tilde{\mathbf{2}}_i,
	&&
	\tilde{\mathbf{3}}_{11} \otimes \tilde{\mathbf{3}}_{21} = \frac{1}{3}\tilde{\mathbf{3}}_{12} + \frac{1}{3} \tilde{\mathbf{3}}_{22},
	&&
	\tilde{\mathbf{3}}_{12} \otimes \tilde{\mathbf{3}}_{22} = \frac{1}{3}\tilde{\mathbf{3}}_{11} + \frac{1}{3} \tilde{\mathbf{3}}_{21},
\end{align}
It shows that $\tilde{\mathbf{3}}_{2i} \otimes \tilde{\mathbf{3}}_{2i}$ does not contain $\tilde{\mathbf{3}}_{2i}^*$ although ${\mathbf{3}}_{2i} \otimes {\mathbf{3}}_{2i}$ does, which is consistent with Eq.~\eqref{eq:projected_phi^3} as expected.

\section{Phenomenological Implications}
\label{sec:pheno}

In the previous section, we have investigated non-invertible selection rules and fusion-like multiplication rules for a non-Abelian symmetry gauged by its subgroup.
We consider their phenomenological implications in this section.

\subsection{Suppressed Couplings}

Before discrete gauging, let us study an $(n+1)$-point coupling in a $G \rtimes H$ symmetric model.
In this model, an $(n+1)$-point coupling is written as 
\begin{align}
	g (\phi_{\alpha_1} \phi_{\alpha_2} \cdots \phi_{\alpha_{n}}\phi_{\alpha_{n+1}}^*)_{\mathbf{1}}
\end{align}
where $g$ is a coupling constant, and $\alpha_i$ denotes an irreducible representation of $G\rtimes H$.
Since $\mathbf{r}^{(\alpha_i)}$ is a unitary representation, the trivial singlet is always obtained as the inner product of two vectors belonging to the same representation.
Using the components of the fields, the $(n+1)$-point coupling can be written as
\begin{align}
	g ({\phi}_{\alpha_1})_i 
	({\phi}_{\alpha_2})_j  \cdots ({\phi}_{\alpha_n})_k  ({\phi}_{\alpha_{n+1}})_l^*
	M^{(\textrm{CG})}_{l, ij\cdots k},
	\label{eq:(n+1)_coupling}
\end{align}
where we introduce the $(n+1)$-point coupling matrix by
\begin{align}
M^{(\textrm{CG})}_{l, ij\cdots k} = \bra{( \alpha_{n+1})_l}\ket{(\alpha_1)_i (\alpha_2)_j \cdots (\alpha_n)_k}.
\end{align}
It is obvious that $(M^{(\textrm{CG})} M^{(\textrm{CG})\dagger})_{lm} = \bra{( \alpha_{n+1})_l}\ket{( \alpha_{n+1})_m} = \delta_{lm}$,
since $\ket{(\alpha_1)_i (\alpha_2)_j \cdots (\alpha_n)_k}$ spans the entire representation space. 
Therefore, the absolute values of the eigenvalues of the coupling matrix are always 1 (or 0 if the interaction is prohibited) for all field components.
Thus, the coupling constants are common to $G\rtimes H$-invariant couplings.
For instance, when $n = 1$, $M^{(\textrm{CG})}_{ij} = \bra{\alpha_i} \ket{\beta_j} = \delta_{\alpha \beta} \delta_{ij}$, and hence the mass is common to all components of the same field.

After gauging, fields are projected onto $H$-invariant subspace, and the  $(n+1)$-bare coupling is also projected as 
\begin{align}
	g (\tilde{\phi}_{\alpha_1})_i 
	(\tilde{\phi}_{\alpha_2})_j  \cdots (\tilde{\phi}_{\alpha_n})_k  (\tilde{\phi}_{\alpha_{n+1}})_l^*
	\tilde M^{(\textrm{CG})} ,
\end{align}
where the projected coupling matrix is given by
\begin{align}
\tilde M^{(\textrm{CG})} = \bra{(\tilde \alpha_{n+1})_l}\ket{(\tilde \alpha_1)_i (\tilde \alpha_2)_j \cdots (\tilde \alpha_n)_k}.
\end{align}
In this case $\ket{(\tilde \alpha_1)_i (\tilde \alpha_2)_j \cdots (\tilde \alpha_n)_k}$ no longer spans the entire representation space,
and the eigenvalues of $\tilde M^{(\textrm{CG})} \tilde M^{(\textrm{CG})\dagger}$ are not necessarily 1.
The absolute values of eigenvalues of $\tilde M^{(\textrm{CG})} \tilde M^{(\textrm{CG}) \dagger}$ are evaluated in the same way of calculating the eigenvalues of a mass matrix.
Since the summation of eigenvalues is given by its trace, we obtain
\begin{align}
	\textrm{Tr}~ \tilde M^{(\textrm{CG})} \tilde M^{(\textrm{CG})^\dagger} = \dim V^{(\alpha_{n+1})} \tilde{N}^{\alpha_{n+1}}_{\alpha_1 \alpha_2 \cdots \alpha_n} 
\end{align}
where we used the relation for the projection-like operator Eq.~\eqref{eq:tilde_N^g_a...a}. 
Since $\tilde M^{(\textrm{CG})} \tilde M^{(\textrm{CG})^\dagger} $ is a matrix acting on $N^{\alpha_{n+1}}_{\alpha_1 \cdots \alpha_n} V^{(\alpha_{n+1})}$, 
there are  $N^{\alpha_{n+1}}_{\alpha_1 \cdots \alpha_n} \dim \tilde V^{(\alpha_{n+1})}$ eigenvalues.
Thus the average of eigenvalues and coupling constants are evaluated as
\begin{align}
	g_{\mathrm{eff}} \sim g \sqrt{\frac{\mathrm{dim} V^{(\alpha_{n+1})}}{\mathrm{dim} \tilde{V}^{(\alpha_{n+1})}}
	\frac{\tilde{N}^{\alpha_{n+1}}_{\alpha_1\alpha_2\cdots \alpha_n}}{{N}^{\alpha_{n+1}}_{\alpha_1\alpha_2\cdots \alpha_n}}}
\end{align}
after gauging.
For example, in $\mathbb{Z}_2$ gauging of $S_4$ model, $\tilde N^{\mathbf{2}}_{\mathbf{3}' \mathbf{3}'} = \frac 1 {12}$ (See Eq.~\eqref{eq:N^g_ab_S}).
Thus the effective 3-point coupling $\tilde \phi_{\mathbf{3}'} \tilde \phi_{\mathbf{3}'} \tilde \phi_{\mathbf{2}}$ is suppressed by a factor of $\sqrt{\frac 21 \frac{1}{12}} = \frac{1}{\sqrt{6}}$.
In general $\tilde{N}^{\alpha_{n+1}}_{\alpha_1 \alpha_2 \cdots \alpha_n}$ takes various values.
Therefore discrete gauging provides an origin of various couplings.

\subsection{Flavor Structure}

The usual non-invertible selection rule determines only whether a coupling is allowed, 
and its value is a free parameter.
However, in the discrete gauging of non-Abelian discrete symmetry, the ratios of the coupling constants among different generations are also fixed, 
strongly constraining the flavor structure.

Here we consider $3$-generation neutrino masses from discrete gauging of non-Abelian symmetry for illustrative purposes.
We assume that there are 3-generations of right-handed neutrinos $\nu_R^i$ and lepton doublets $L^i$, and one Higgs doublet $\Phi$.
We consider a model with $\mathbb{Z}_2$ gauging of $\Delta(54)$ global symmetry.
We assume $\nu_R^i$ transform as $({\mathbf{3}}_{11}, {\mathbf{3}}_{21})$, $L^i$ transform as $({\mathbf{3}}_{11}, {\mathbf{3}}_{21})$, and $\Phi$ transforms as ${\mathbf{2}}_2$ under $\Delta(54)$.
Before gauging, we have 6-generations of neutrinos and 2 Higgs doublets, but $\mathbb{Z}_2$-gauging project out half of the fields, and hence a three-generation lepton sector is realized.
The bare Yukawa interactions are given by  
\begin{align}
	\Phi \bar{\nu}_R^i Y_{ij} L^j  + (h.c.) 
	= \Phi
	\begin{pmatrix}
	\bar \nu_R^1 & \bar \nu_R^2 & \bar \nu_R^3
	\end{pmatrix}
	\begin{pmatrix}
	\frac{\omega}{\sqrt 6} g_1& \frac{\omega}{\sqrt 3} g_1& \frac{\omega}{\sqrt 6}g_2\\
	\frac{\omega^2}{\sqrt 3} g_1& 0 & -\frac{\omega^2}{\sqrt 3}g_2\\
	-\frac{1}{\sqrt 6} g_3& \frac{\omega^2}{\sqrt 3} g_3& -\frac{\omega}{\sqrt 6}g_4\\
	\end{pmatrix}
	\begin{pmatrix}
	L^1 \\
	L^2 \\
	L^3
	\end{pmatrix} + (h.c.).
\end{align}
$\Phi \bar{\nu}_R^2 L^2$ is prohibited since Table \ref{tab:Delta54_2} shows that $(\tilde{\mathbf{3}}_{11})_2 \otimes (\tilde{\mathbf{3}}_{12})_2 $ does not contain $\tilde{\mathbf{2}}_2$.
$g_{1,2,3,4}$ denotes coupling constants.
$\tilde{\mathbf{3}}_{11}$ contains two independent fields, there are only 4 independent coupling constants 
in contrast to the Abelian case.
The couplings between the first and second generations are constrained by the CG coefficients.

When all $g_i$ have approximately equal values, 
the mass matrix is approximated by 
\begin{align}
	M \sim 
	\begin{pmatrix}
	\frac{\omega}{\sqrt 6}& \frac{\omega}{\sqrt 3} & \frac{\omega}{\sqrt 6}\\
	\frac{\omega^2}{\sqrt 3} & 0 & -\frac{\omega^2}{\sqrt 3}\\
	-\frac{1}{\sqrt 6}& \frac{\omega^2}{\sqrt 3}& -\frac{\omega}{\sqrt 6}\\
	\end{pmatrix}.
\end{align}
When the charged lepton mass matrix is already diagonalized, lepton mixing angles solely come from the neutrino mass matrix, 
and large mixing angles are naturally obtained.
Since loop corrections violate the non-invertible selection rules,
it is also interesting to investigate the loop corrections of non-invertible selection rule from discrete gauging of non-Abelian  symmetry \cite{Heckman:2024obe, Kaidi:2024wio, Funakoshi:2024uvy, Dong:2026crl, Suzuki:2025bxg, Suzuki:2025kxz, Xu:2026nwh}, 
but it is beyond our scope in this paper, and we will study it elsewhere.

\section{Conclusion}
\label{sec:conclusion}

We have investigated the discrete $H$-gauging of theories with an underlying global symmetry group $G$. 
While previous works have exclusively focused on models where $G$ is Abelian, 
we have systematically extended the analysis to non-Abelian groups $G$ with generally non-Abelian $H$-gauging. 
When $G$ is Abelian and each singlet representation in $G$ corresponds to an $H$-orbit of conjugacy classes of $G$, 
bare couplings are allowed simply if the sum of the $G$-charges carried by the constituent fields cancels.
For general $G$ and $H$, however, there is no obvious one-to-one correspondence 
between irreducible representations and $H$-orbits of conjugacy classes. 
Moreover, when $G$ is non-Abelian, its irreducible representations are essentially multidimensional. 
Consequently, the group transformations induced by $H$ not only permute distinct $G$-representations 
but also non-trivially mix the internal components within a single $G$-multiplet, potentially projecting out specific degrees of freedom. 
As a result, the conventional coupling selection rules based on standard tensor product decompositions 
are no longer sufficient to determine the non-zero interactions. 
To accurately determine the remaining physical states and their allowed bare couplings, 
we formulated a rigorous theoretical framework for $H$-gauged models incorporating generalized field transformations. 
Within this framework, we analyze the representation spaces of the full semidirect product $G\rtimes H$ 
and explicitly evaluate how the elements of $H$ transform the individual vector components.

From the viewpoint of representations of $G\rtimes H$, 
the transformations induced by $H$ are closed within each irreducible representation. 
The $H$-gauging procedure projects out a portion of the fields contained in these irreducible representations. 
Due to $H$-gauging, only a restricted subset of the fields remains, and the standard tensor product algebra is broken. 
However, by introducing the projected characters, 
we showed that the characters restricted to the remaining subspace still hold useful properties, including orthogonality. 
This allows us to systematically determine which representations appear in the products. 
Using these projected characters, we derived the necessary and sufficient conditions 
for non-vanishing $n$-point couplings in an elegant and compact form, 
analogous to that of a conventional group symmetry. 
Furthermore, we demonstrated that while the $H$-invariant vector spaces themselves 
do not form a closed algebra, their individual components obey a well-defined, associative fusion-like algebra 
governed by the Clebsch-Gordan (CG) coefficients for the $H$-invariant fields. 
By using this fusion-like algebra recursively, we can alternatively derive the non-invertible selection rules. 
The results obtained from these two distinct formulas are indeed consistent.

This framework opens up interesting phenomenological applications. 
In particular, these rules not only constrain whether a bare coupling is allowed, 
but also give specific relations among the allowed coupling constants of multiplet interactions, 
whose ratios are characterized by the CG coefficients for $H$-invariant fields.
This feature distinguishes the present construction from earlier non-invertible selection rules 
based solely on a simple fusion algebra among conjugacy classes. 
Its application to particle phenomenology is therefore interesting, particularly in the context of lepton flavor models. 
On the other hand, these constraints are generally expected to be violated by quantum corrections~\cite{
%Heckman:2024obe, 
Kaidi:2024wio, Funakoshi:2024uvy, Dong:2026crl, Suzuki:2025bxg, Suzuki:2025kxz, Xu:2026nwh}. 
Applying this framework to flavor physics, as well as evaluating how such loop effects would affect these constraints, 
remain important issues for future investigation.

As for a more theoretical direction, we utilized CG coefficients 
to construct an associative algebra that yields the non-invertible selection rules for theories with discrete gauging. 
This suggests the existence of an associativity isomorphism $a_{\alpha,\beta,\gamma}: (\tilde{\mathbf{r}}^{(\alpha)} \otimes \tilde{\mathbf{r}}^{(\beta)})\otimes \tilde{\mathbf{r}}^{(\gamma)} \to \tilde{\mathbf{r}}^{(\alpha)} \otimes ( \tilde{\mathbf{r}}^{(\beta)}\otimes \tilde{\mathbf{r}}^{(\gamma)})$, allowing one to formulate this structure in terms of a monoidal category or a fusion category, 
rather than a simple associative algebra~\cite{Etingof:2015, Thorngren:2021yso}. 
Exploring this categorical structure is an interesting theoretical direction that we leave for future work.

\section*{Acknowledgements}
H.~O.~is supported in part by JSPS KAKENHI Grants No. 21K03554 and No. 22H00138.
The authors thank Tatsuo Kobayashi and Ryusei Nishida for helpful comments about violation of associativity and fusion categories.

\appendix

\section{Orthogonality Relations for $H$-invariant Subspaces}
\label{sec:tilde_rho}

Here, we study transformations restricted to $H$-invariant subspaces and their associated characters.
Let $\tilde \rho^{(\alpha)}$ denote the projected matrices associated with the projection operator onto the $H$-invariant subspace.
We find
\begin{align}
	\frac 1 {|G \rtimes H|} \sum_{g \in G \rtimes H} \tilde{\rho}_{ji}^{(\alpha)}(g^{-1}) \tilde{\rho}_{kl}^{(\beta)}(g)
	&= \frac 1 {|G \rtimes H|}\sum_{g \in G \rtimes H} \sum_{p,q,r,s} P^{(\alpha)}_{jp} {\rho}_{pq}^{(\alpha)}(g^{-1}) P^{(\alpha)}_{qi} P^{(\beta)}_{kr}  {\rho}_{rs}^{(\beta)}(g) P^{(\beta)}_{sl}
	\notag
	\\
	&=\frac1{\mathrm{dim}\,V^{(\alpha)}} \delta_{\alpha\beta} P^{(\beta)}_{ki} P^{(\alpha)}_{jl},
\end{align}
where we use a orthogonality relation among irreducible representations of $G\rtimes H$: 
\begin{align}
	\frac{1}{|G \rtimes H|}\sum_{g \in G \rtimes H} \rho^{(\alpha)}_{pq}(g^{-1}) \rho^{(\beta)}_{rs}(g) = \frac{1}{\mathrm{dim}\,V^{(\alpha)}} \delta_{ps}\delta_{qr}\delta_{\alpha \beta}.
\end{align}
Taking trace of both sides, we obtain
\begin{align}
	\frac{1}{|G \rtimes H|} \sum_{g\in G \rtimes H} \tilde{\chi}^{(\alpha)*}(g) \tilde{\chi}^{(\beta)}(g) 
	= \frac{1}{\mathrm{dim}\,V^{(\alpha)}} \delta_{\alpha \beta} \tr P^{(\alpha)}.
\end{align}
It is always possible to diagonalize $P^{(\alpha)}$, and hence we obtain
\begin{align}
	\frac 1 {|G \rtimes H|} \sum_{g\in G \rtimes H} \tilde{\chi}^{(\alpha)*}(g) \tilde{\chi}^{(\beta)}(g) =  \frac{\mathrm{dim}\, \tilde V^{(\alpha)}}{\mathrm{dim}\, V^{(\alpha)}} \delta_{\alpha \beta}.
	\label{eq:auth_orbifold_rep}
\end{align}
This relation is generalized to including $\tilde \rho^{(\alpha)}_\perp$, which is a projected matrix on the complemental space of $V^{(\alpha)}$.
We find 
\begin{align}
	\frac 1 {|G \rtimes H|} \sum_{g \in G \rtimes H} (\tilde{\rho}_\perp)^{(\alpha)}_{ji}(g^{-1}) \tilde{\rho}_{kl}^{(\beta)}(g)
	&=\frac1{\mathrm{dim}\, V^{(\alpha)}} \delta_{\alpha\beta} (P^{(\beta)}P_\perp^{(\alpha)})_{ki} (P_\perp^{(\alpha)}P^{(\beta)})_{jl}
	= 0,
\end{align}
and $\tilde \rho^{(\alpha)}$ and $\tilde \rho^{(\alpha)}_\perp$ are orthogonal.
Thus we obtain the following relations
\begin{align}
	\frac 1 {|G \rtimes H|} \sum_{g\in G \rtimes H} \tilde {\chi}^{(\alpha)*}(g) \tilde {\chi}^{(\beta)}(g) &=  
	\frac{\mathrm{dim}\,\tilde {V}^{(\alpha)}}{\mathrm{dim}\,{V}^{(\alpha)}} \delta_{\alpha \beta},
	\quad
	\frac 1 {|G \rtimes H|} \sum_{g\in G \rtimes H} \tilde {\chi}^{(\alpha)*}(g) \tilde {\chi}_\perp^{(\beta)}(g) =  
	0,
	\notag
	\\
	\frac 1 {|G \rtimes H|} \sum_{g\in G \rtimes H} \tilde {\chi}_\perp^{(\alpha)*}(g) \tilde {\chi}_\perp^{(\beta)}(g) &=  
	\frac{\mathrm{dim}\,\tilde {V}_\perp^{(\alpha)}}{\mathrm{dim}\,{V}^{(\alpha)}} \delta_{\alpha \beta}.
\end{align}
$\tilde \chi^{(\alpha)}$ and $\tilde \chi^{(\alpha)}_\perp$ also holds orthogonality relation.
This relation shows that the gauge invariant characters are also orthogonal each other
and it properly counts the number of remaining fields after discrete gauging.

\section{Clebsch-Gordan Coefficients for $\mathbb{Z}_2$-gauged Fields in the $\Delta(54)$ Model}
\label{sec:CG_D54}

We summarize the multiplication rules for the $\mathbb{Z}_2$-gauged fields in the $\Delta(54)$ model.
We use the  representation basis defined in  \cite{Kobayashi:2022moq}.
Since $\Delta(54)$ has a non-trivial center isomorphic to $\mathbb{Z}_3$, the fields carry $\mathbb{Z}_3$ charges.
Because the gauged $\mathbb{Z}_2$ commutes with this center, the multiplication rules conserve these $\mathbb{Z}_3$ charges.

\begin{landscape}

\begin{table}[htb!]
\begin{center}
\renewcommand{\arraystretch}{1.35} 
\setlength{\tabcolsep}{5pt} 
{\small
\begin{tabular}{| c | c | c | c | c  |}
\hline
& $\mathbb{Z}_3$
& $(\tilde \phi_{\mathbf{3}_{11}})_1$  
& $(\tilde \phi_{\mathbf{3}_{11}})_2$  
& $\tilde \phi_{\mathbf{3}_{21}}$ 
 \\ 
\hline

$(\tilde \phi_{\mathbf{3}_{11}})_1$ 
& $\omega$
& $\frac 1 {\sqrt 2} (\tilde \phi_{\mathbf{3}_{12}})_1 + \frac 1 {\sqrt 2} (\tilde \phi_{\mathbf{3}_{12}})_2$  
& $\frac 1 {\sqrt 2} (\tilde \phi_{\mathbf{3}_{12}})_1 - \frac 1 {\sqrt 2} \tilde \phi_{\mathbf{3}_{22}}$ 
& $\frac 1 {\sqrt 2} \tilde \phi_{\mathbf{3}_{22}} - \frac 1 {\sqrt 2} (\tilde \phi_{\mathbf{3}_{12}})_2$ 
\\

$(\tilde \phi_{\mathbf{3}_{11}})_2$ 
& $\omega$
& $\frac 1 {\sqrt 2} (\tilde \phi_{\mathbf{3}_{12}})_1 + \frac 1 {\sqrt 2} \tilde \phi_{\mathbf{3}_{22}}$
& $(\tilde \phi_{\mathbf{3}_{12}})_2$ 
& $\frac 1 {\sqrt 2} (\tilde \phi_{\mathbf{3}_{12}})_1 - \frac 1 {\sqrt 2} \tilde \phi_{\mathbf{3}_{22}}$ 
\\

$\tilde \phi_{\mathbf{3}_{21}}$ 
& $\omega$
& $\frac 1 {\sqrt 2} \tilde \phi_{\mathbf{3}_{22}} - \frac 1 {\sqrt 2} (\tilde \phi_{\mathbf{3}_{12}})_2$ 
& $\frac 1 {\sqrt 2} (\tilde \phi_{\mathbf{3}_{12}})_1 - \frac 1 {\sqrt 2} \tilde \phi_{\mathbf{3}_{22}}$ 
& $\frac 1{\sqrt 2} (\tilde \phi_{\mathbf{3}_{12}})_1 - \frac 1{\sqrt 2} (\tilde \phi_{\mathbf{3}_{12}})_2$
\\

\hline

$(\tilde \phi_{\mathbf{3}_{12}})_1$ 
& $\omega^2$
& $\frac 1{\sqrt 3} \tilde \phi_{\mathbf{1}} - \frac{\omega^2}{\sqrt 6} \tilde \phi_{\mathbf{2}_1} + \frac{\omega}{\sqrt 6} \tilde \phi_{\mathbf{2}_2} + \frac{1}{\sqrt 6} \tilde \phi_{\mathbf{2}_3} + \frac{1}{\sqrt 6} \tilde \phi_{\mathbf{2}_4}$
& $\frac{\omega^2}{\sqrt 3} \tilde \phi_{\mathbf{2}_2} + \frac{\omega^2}{\sqrt 3} \tilde \phi_{\mathbf{2}_3} +  \frac{1}{\sqrt 3} \tilde \phi_{\mathbf{2}_4}$ 
& $\frac{i\omega^2}{\sqrt 2} \tilde \phi_{\mathbf{2}_1} - \frac 1 {\sqrt 6} \tilde \phi_{\mathbf{2}_2} + \frac {\omega} {\sqrt 6} \tilde \phi_{\mathbf{2}_3} + \frac 1{\sqrt 6} \tilde \phi_{\mathbf{2}_4}$
\\

$(\tilde \phi_{\mathbf{3}_{12}})_2$ 
& $\omega^2$
& $\frac{1}{\sqrt 3} \tilde \phi_{\mathbf{2}_2} + \frac{\omega}{\sqrt 3} \tilde \phi_{\mathbf{2}_3} +  \frac{1}{\sqrt 3} \tilde \phi_{\mathbf{2}_4}$ 
& $\frac{1}{\sqrt 3} \tilde \phi_{\mathbf{1}} + \omega^2 \sqrt{\frac 23} \tilde \phi_{\mathbf{2}_1}$
& $\frac{\omega^2}{\sqrt 3} \tilde \phi_{\mathbf{2}_2} - \frac {\omega^2} {\sqrt 3} \tilde \phi_{\mathbf{2}_3} - \frac 1 {\sqrt 3} \tilde \phi_{\mathbf{2}_4} $
\\

$\tilde \phi_{\mathbf{3}_{22}}$ 
& $\omega^2$
& $-\frac{i\omega}{\sqrt 2} \tilde \phi_{\mathbf{2}_1} + \frac \omega{\sqrt 6} \tilde \phi_{\mathbf{2}_2} - \frac 1 {\sqrt 6} \tilde \phi_{\mathbf{2}_3} -\frac 1{\sqrt 6} \tilde \phi_{\mathbf{2}_4}$
& $-\frac{\omega^2}{\sqrt 3} \tilde \phi_{\mathbf{2}_2} + \frac {\omega^2} {\sqrt 3} \tilde \phi_{\mathbf{2}_3} + \frac 1 {\sqrt 3} \tilde \phi_{\mathbf{2}_4} $
& $\frac{1}{\sqrt 3} \tilde \phi_{\mathbf{1}} - \frac{\omega^2}{\sqrt 6} \tilde \phi_{\mathbf{2}_1} - \frac{\omega}{\sqrt 6} \tilde \phi_{\mathbf{2}_2} - \frac{\omega^2}{\sqrt 6} \tilde \phi_{\mathbf{2}_3}- \frac{1}{\sqrt 6} \tilde \phi_{\mathbf{2}_4}$
\\

\hline
\end{tabular}
}
\end{center}
\vspace{-5mm} \caption{Multiplication rules for $\mathbb{Z}_2$-gauging of $\Delta(54)$ fields. }
\label{tab:Delta54_2}
\end{table}

\begin{table}[htb!]
\begin{center}
\renewcommand{\arraystretch}{1.35} 
\setlength{\tabcolsep}{5pt} 
{\small
\begin{tabular}{| c | c | c | c | c |}
\hline
& $\mathbb{Z}_3$
& $(\tilde \phi_{\mathbf{3}_{12}})_1$  
& $(\tilde \phi_{\mathbf{3}_{12}})_2$  
& $\tilde \phi_{\mathbf{3}_{22}}$ 
 \\ 
\hline

$(\tilde \phi_{\mathbf{3}_{11}})_1$ 
& $\omega$
& $\frac 1{\sqrt 3} \tilde \phi_{\mathbf{1}} - \frac{\omega^2}{\sqrt 6} \tilde \phi_{\mathbf{2}_1} + \frac{\omega}{\sqrt 6} \tilde \phi_{\mathbf{2}_2} + \frac{1}{\sqrt 6} \tilde \phi_{\mathbf{2}_3} + \frac{1}{\sqrt 6} \tilde \phi_{\mathbf{2}_4}$
& $\frac{1}{\sqrt 3} \tilde \phi_{\mathbf{2}_2} + \frac{\omega}{\sqrt 3} \tilde \phi_{\mathbf{2}_3} +  \frac{1}{\sqrt 3} \tilde \phi_{\mathbf{2}_4}$  
& $-\frac{i\omega}{\sqrt 2} \tilde \phi_{\mathbf{2}_1} + \frac \omega{\sqrt 6} \tilde \phi_{\mathbf{2}_2} - \frac 1 {\sqrt 6} \tilde \phi_{\mathbf{2}_3} -\frac 1{\sqrt 6} \tilde \phi_{\mathbf{2}_4}$
\\

$(\tilde \phi_{\mathbf{3}_{11}})_2$ 
& $\omega$
& $\frac{\omega^2}{\sqrt 3} \tilde \phi_{\mathbf{2}_2} + \frac{\omega^2}{\sqrt 3} \tilde \phi_{\mathbf{2}_3} +  \frac{1}{\sqrt 3} \tilde \phi_{\mathbf{2}_4}$ 
& $\frac{1}{\sqrt 3} \tilde \phi_{\mathbf{1}} + \omega^2 \sqrt{\frac 23} \tilde \phi_{\mathbf{2}_1}$
& $-\frac{\omega^2}{\sqrt 3} \tilde \phi_{\mathbf{2}_2} + \frac {\omega^2} {\sqrt 3} \tilde \phi_{\mathbf{2}_3} + \frac 1 {\sqrt 3} \tilde \phi_{\mathbf{2}_4} $
\\

$\tilde \phi_{\mathbf{3}_{21}}$ 
& $\omega$
& $\frac{i\omega^2}{\sqrt 2} \tilde \phi_{\mathbf{2}_1} - \frac 1 {\sqrt 6} \tilde \phi_{\mathbf{2}_2} + \frac {\omega} {\sqrt 6} \tilde \phi_{\mathbf{2}_3} + \frac 1{\sqrt 6} \tilde \phi_{\mathbf{2}_4}$
& $\frac{\omega^2}{\sqrt 3} \tilde \phi_{\mathbf{2}_2} - \frac {\omega^2} {\sqrt 3} \tilde \phi_{\mathbf{2}_3} - \frac 1 {\sqrt 3} \tilde \phi_{\mathbf{2}_4} $
& $\frac{1}{\sqrt 3} \tilde \phi_{\mathbf{1}} - \frac{\omega^2}{\sqrt 6} \tilde \phi_{\mathbf{2}_1} - \frac{\omega}{\sqrt 6} \tilde \phi_{\mathbf{2}_2} - \frac{\omega^2}{\sqrt 6} \tilde \phi_{\mathbf{2}_3}- \frac{1}{\sqrt 6} \tilde \phi_{\mathbf{2}_4}$
\\

\hline

$(\tilde \phi_{\mathbf{3}_{12}})_1$ 
& $\omega^2$
& $\frac 1 {\sqrt 2} (\tilde \phi_{\mathbf{3}_{11}})_1 + \frac 1 {\sqrt 2} (\tilde \phi_{\mathbf{3}_{11}})_2$  
& $\frac 1 {\sqrt 2} (\tilde \phi_{\mathbf{3}_{11}})_1 - \frac 1 {\sqrt 2} \tilde \phi_{\mathbf{3}_{21}}$ 
& $\frac 1 {\sqrt 2} \tilde \phi_{\mathbf{3}_{21}} - \frac 1 {\sqrt 2} (\tilde \phi_{\mathbf{3}_{11}})_2$
\\

$(\tilde \phi_{\mathbf{3}_{12}})_2$ 
& $\omega^2$
& $\frac 1 {\sqrt 2} (\tilde \phi_{\mathbf{3}_{11}})_1 +  \frac 1 {\sqrt 2} \tilde \phi_{\mathbf{3}_{21}}$  
& $(\tilde \phi_{\mathbf{3}_{11}})_2$ 
& $\frac 1 {\sqrt 2} (\tilde \phi_{\mathbf{3}_{11}})_1 - \frac 1 {\sqrt 2} \tilde \phi_{\mathbf{3}_{21}}$
\\

$\tilde \phi_{\mathbf{3}_{22}}$ 
& $\omega^2$
& $\frac 1 {\sqrt 2} \tilde \phi_{\mathbf{3}_{21}} - \frac 1 {\sqrt 2} (\tilde \phi_{\mathbf{3}_{11}})_2$
& $\frac 1 {\sqrt 2} (\tilde \phi_{\mathbf{3}_{11}})_1 - \frac 1 {\sqrt 2} \tilde \phi_{\mathbf{3}_{21}}$
& $\frac 1 {\sqrt 2}(\tilde \phi_{\mathbf{3}_{11}})_1 - \frac 1 {\sqrt 2}(\tilde \phi_{\mathbf{3}_{11}})_2$
\\

\hline
\end{tabular}
}
\end{center}
\vspace{-5mm} \caption{Multiplication rules for $\mathbb{Z}_2$-gauging of $\Delta(54)$ fields. }
\label{tab:Delta54_3}
\end{table}

\begin{table}[htb!]
\begin{center}
\renewcommand{\arraystretch}{1.35} 
\setlength{\tabcolsep}{5pt} 
{\small
\begin{tabular}{| c | c | c | c | c  | c | c|}
\hline
& $\mathbb{Z}_3$
& $\tilde \phi_{\mathbf{1}}$  
& $\tilde \phi_{\mathbf{2}_1}$  
& $ \tilde \phi_{\mathbf{2}_2}$ 
& $\tilde \phi_{\mathbf{2}_3} $ 
& $\tilde \phi_{\mathbf{2}_4} $ 
 \\ 
\hline

$\tilde \phi_{\mathbf{1}}$ 
& $1$
& $\tilde \phi_{\mathbf{1}}$  
& $\tilde \phi_{\mathbf{2}_1}$  
& $ \tilde \phi_{\mathbf{2}_2}$ 
& $\tilde \phi_{\mathbf{2}_3} $ 
& $\tilde \phi_{\mathbf{2}_4} $ \\ 

$\tilde \phi_{\mathbf{2}_1}$ 
& $1$
& $\tilde \phi_{\mathbf{2}_1}$  
& $\frac 1{\sqrt 2}  \tilde \phi_{\mathbf{1}} +  \frac 1{\sqrt 2} \tilde \phi_{\mathbf{2}_1}$ 
& $\frac 1{\sqrt 2}  \tilde \phi_{\mathbf{2}_3} +  \frac 1{\sqrt 2} \tilde \phi_{\mathbf{2}_4}$ 
& $\frac 1{\sqrt 2}  \tilde \phi_{\mathbf{2}_2} +  \frac 1{\sqrt 2} \tilde \phi_{\mathbf{2}_4}$  
& $\frac 1{\sqrt 2}  \tilde \phi_{\mathbf{2}_2} +  \frac 1{\sqrt 2} \tilde \phi_{\mathbf{2}_3}$\\

$\tilde \phi_{\mathbf{2}_2}$ 
& $1$
& $\tilde \phi_{\mathbf{2}_2}$  
& $\frac 1{\sqrt 2}  \tilde \phi_{\mathbf{2}_3} +  \frac 1{\sqrt 2} \tilde \phi_{\mathbf{2}_4}$ 
& $\frac 1{\sqrt 2}  \tilde \phi_{\mathbf{1}} +  \frac 1{\sqrt 2} \tilde \phi_{\mathbf{2}_2}$ 
& $\frac 1{\sqrt 2}  \tilde \phi_{\mathbf{2}_1} +  \frac 1{\sqrt 2} \tilde \phi_{\mathbf{2}_4}$  
& $\frac 1{\sqrt 2}  \tilde \phi_{\mathbf{2}_1} +  \frac 1{\sqrt 2} \tilde \phi_{\mathbf{2}_3}$\\

$\tilde \phi_{\mathbf{2}_3}$ 
& $1$
& $\tilde \phi_{\mathbf{2}_3}$  
& $\frac 1{\sqrt 2}  \tilde \phi_{\mathbf{2}_2} +  \frac 1{\sqrt 2} \tilde \phi_{\mathbf{2}_4}$ 
& $\frac 1{\sqrt 2}  \tilde \phi_{\mathbf{2}_1} +  \frac 1{\sqrt 2} \tilde \phi_{\mathbf{2}_4}$ 
& $\frac 1{\sqrt 2}  \tilde \phi_{\mathbf{1}} +  \frac 1{\sqrt 2} \tilde \phi_{\mathbf{2}_3}$  
& $\frac 1{\sqrt 2}  \tilde \phi_{\mathbf{2}_1} +  \frac 1{\sqrt 2} \tilde \phi_{\mathbf{2}_2}$\\

$\tilde \phi_{\mathbf{2}_4}$ 
& $1$
& $\tilde \phi_{\mathbf{2}_4}$  
& $\frac 1{\sqrt 2}  \tilde \phi_{\mathbf{2}_2} +  \frac 1{\sqrt 2} \tilde \phi_{\mathbf{2}_3}$ 
& $\frac 1{\sqrt 2}  \tilde \phi_{\mathbf{2}_1} +  \frac 1{\sqrt 2} \tilde \phi_{\mathbf{2}_3}$ 
& $\frac 1{\sqrt 2}  \tilde \phi_{\mathbf{2}_1} +  \frac 1{\sqrt 2} \tilde \phi_{\mathbf{2}_2}$  
& $\frac 1{\sqrt 2}  \tilde \phi_{\mathbf{1}} +  \frac 1{\sqrt 2} \tilde \phi_{\mathbf{2}_4}$\\

\hline

$(\tilde \phi_{\mathbf{3}_{11}})_1$ 
& $\omega$
& $(\tilde \phi_{\mathbf{3}_{11}})_1$  
& $- \frac 1 {2} \omega  (\tilde \phi_{\mathbf{3}_{11}})_1 -  i \omega \frac { \sqrt 3}{2} \tilde \phi_{\mathbf{3}_{21}}$ 
& $\frac 1 {2}  (\tilde \phi_{\mathbf{3}_{11}})_1 +\frac1{\sqrt 2}  \omega^2 (\tilde \phi_{\mathbf{3}_{11}})_2 - \frac 12 \tilde \phi_{\mathbf{3}_{21}}$
& $\frac 1 {2}  \omega (\tilde \phi_{\mathbf{3}_{11}})_1 + \frac1{\sqrt 2}  \omega^2 (\tilde \phi_{\mathbf{3}_{11}})_2 + \frac 12 \omega \tilde \phi_{\mathbf{3}_{21}}$
& $\frac 1 {2} \omega  (\tilde \phi_{\mathbf{3}_{11}})_1 + \frac {1}{\sqrt 2} (\tilde \phi_{\mathbf{3}_{11}})_2 + \frac 12 \tilde \phi_{\mathbf{3}_{21}}$ 
\\

$(\tilde \phi_{\mathbf{3}_{11}})_2$ 
& $\omega$
& $(\tilde \phi_{\mathbf{3}_{11}})_2$  
& $\omega (\tilde \phi_{\mathbf{3}_{11}})_2$ 
& $\frac 1 {\sqrt 2} \omega (\tilde \phi_{\mathbf{3}_{11}})_1 +\frac1{\sqrt 2}  \omega \tilde \phi_{\mathbf{3}_{21}}$
& $\frac 1 {\sqrt 2} (\tilde \phi_{\mathbf{3}_{11}})_1 - \frac1{\sqrt 2}  \tilde \phi_{\mathbf{3}_{21}}$ 
& $\frac 1 {\sqrt 2} (\tilde \phi_{\mathbf{3}_{11}})_1 - \frac1{\sqrt 2}  \tilde \phi_{\mathbf{3}_{21}}$ 
\\

$\tilde \phi_{\mathbf{3}_{21}}$ 
& $\omega$
& $\tilde \phi_{\mathbf{3}_{21}}$  
& $- \frac12 \omega \tilde \phi_{\mathbf{3}_{21}} - i \omega \frac{\sqrt 3}{2} (\tilde \phi_{\mathbf{3}_{11}})_1$ 
& $ \frac 1 {2}  (\tilde \phi_{\mathbf{3}_{11}})_1 - \frac1{\sqrt 2}  \omega^2 (\tilde \phi_{\mathbf{3}_{11}})_2 - \frac 12 \tilde \phi_{\mathbf{3}_{21}}$
& $- \frac 1 {2}  \omega (\tilde \phi_{\mathbf{3}_{11}})_1 + \frac1{\sqrt 2}  \omega^2 (\tilde \phi_{\mathbf{3}_{11}})_2 - \frac 12 \omega \tilde \phi_{\mathbf{3}_{21}}$
& $ - \frac 1 {2}  (\tilde \phi_{\mathbf{3}_{11}})_1 + \frac1{\sqrt 2} (\tilde \phi_{\mathbf{3}_{11}})_2 - \frac 12 \tilde \phi_{\mathbf{3}_{21}}$
\\

\hline

$(\tilde \phi_{\mathbf{3}_{12}})_1$ 
& $\omega^2$
& $(\tilde \phi_{\mathbf{3}_{12}})_1$  
& $- \frac 1 {2} \omega  (\tilde \phi_{\mathbf{3}_{12}})_1 -  i \omega \frac { \sqrt 3}{2} \tilde \phi_{\mathbf{3}_{22}}$ 
& $\frac 1 {2}  \omega (\tilde \phi_{\mathbf{3}_{12}})_1 +\frac1{\sqrt 2}  \omega^2 (\tilde \phi_{\mathbf{3}_{12}})_2 + \frac 12 \omega \tilde \phi_{\mathbf{3}_{22}}$
& $\frac 1 {2}  (\tilde \phi_{\mathbf{3}_{12}})_1 + \frac1{\sqrt 2}  \omega^2 (\tilde \phi_{\mathbf{3}_{12}})_2 - \frac 12 \tilde \phi_{\mathbf{3}_{22}}$
& $\frac 1 {2} \omega  (\tilde \phi_{\mathbf{3}_{12}})_1 + \frac {1}{\sqrt 2} (\tilde \phi_{\mathbf{3}_{12}})_2 - \frac 12 \tilde{\phi}_{\mathbf{3}_{22}}$ 
\\

$(\tilde \phi_{\mathbf{3}_{12}})_2$ 
& $\omega^2$
& $(\tilde \phi_{\mathbf{3}_{12}})_2$  
& $\omega (\tilde \phi_{\mathbf{3}_{12}})_2$ 
& $\frac 1 {\sqrt 2} (\tilde \phi_{\mathbf{3}_{12}})_1 - \frac1{\sqrt 2} \tilde \phi_{\mathbf{3}_{22}}$
& $\frac 1 {\sqrt 2} \omega (\tilde \phi_{\mathbf{3}_{12}})_1 + \frac1{\sqrt 2}  \omega \tilde \phi_{\mathbf{3}_{22}}$ 
& $\frac 1 {\sqrt 2} (\tilde \phi_{\mathbf{3}_{12}})_1 + \frac1{\sqrt 2}  \tilde \phi_{\mathbf{3}_{22}}$ 
\\

$\tilde \phi_{\mathbf{3}_{22}}$ 
& $\omega^2$
& $\tilde \phi_{\mathbf{3}_{22}}$  
& $- \frac12 \omega \tilde \phi_{\mathbf{3}_{22}} - i \omega \frac{\sqrt 3}{2} (\tilde \phi_{\mathbf{3}_{12}})_1$ 
& $ - \frac 1 {2}  \omega (\tilde \phi_{\mathbf{3}_{12}})_1 + \frac1{\sqrt 2}  \omega^2 (\tilde \phi_{\mathbf{3}_{12}})_2 - \frac 12 \omega \tilde \phi_{\mathbf{3}_{22}}$
& $ \frac 1 {2}  (\tilde \phi_{\mathbf{3}_{12}})_1 - \frac1{\sqrt 2}  \omega^2 (\tilde \phi_{\mathbf{3}_{12}})_2 - \frac 12 \omega \tilde \phi_{\mathbf{3}_{22}}$
& $ \frac 1 {2}  (\tilde \phi_{\mathbf{3}_{12}})_1 - \frac1{\sqrt 2} (\tilde \phi_{\mathbf{3}_{12}})_2 - \frac 12 \tilde \phi_{\mathbf{3}_{22}}$
\\

\hline
\end{tabular}
}
\end{center}
\vspace{-5mm} \caption{Multiplication rules for $\mathbb{Z}_2$-gauging of $\Delta(54)$ fields. }
\label{tab:Delta54_1}
\end{table}

\end{landscape}

\bibliography{draft.bib} 
\bibliographystyle{JHEP}

\end{document}